\documentclass[%
 reprint,
nofootinbib,
 amsmath,amssymb,
 aps,
]{revtex4-1}
\usepackage{braket}
\usepackage{amsmath}
\usepackage{mathtools}
\usepackage{hyperref}
\usepackage{graphicx}
\usepackage{dcolumn}
\usepackage{xcolor}
\usepackage{soul}
\usepackage{nicefrac}
\usepackage{bm}

\begin{document}

\preprint{APS/123-QED}

\title{Quantum stochastic trajectories for particles and fields \\ based on positive P-representation}

\author{Stasis Chuchurka}%
\email{stasis.chuchurka@desy.de}
\affiliation{
Deutsches Elektronen-Synchrotron DESY, Hamburg 22607, Germany\;
\\
Department of Physics, Universität Hamburg, Hamburg 22761, Germany
}%

\author{Andrei Benediktovitch}
\affiliation{
Deutsches Elektronen-Synchrotron DESY, Hamburg 22607, Germany
}%

\author{Nina Rohringer}
\email{nina.rohringer@desy.de}
\affiliation{
CFEL, Deutsches Elektronen-Synchrotron DESY, Hamburg 22607, Germany\,
\\
I. Institut für Theoretische Physik, Universität Hamburg, Hamburg 22761, Germany
}%

\begin{abstract}
[...]
\end{abstract}







\begin{abstract}
In this work we introduce a phase-space description based on the positive P representation for bosonic fields interacting with a system of quantum emitters. The formalism is applicable to collective light-matter interactions and open quantum systems with decoherence. Conservation of particle numbers is considered, and a Jordan-Schwinger transformation enables the representation of multi-level quantum emitters. The evolution of the phase-space description of the combined system of emitters and field is formulated in terms of stochastic trajectories and we derive the rules of mapping from traditional quantum mechanics to this stochastic formalism. The resulting equations of motion encode deterministic, classical evolution with quantum effects incorporated by stochastic noise terms. The framework's equations and properties are provided without specifying the Hamiltonian, aiming for broad applicability in diverse research domains. A potential future application is the quantum mechanical description of collective spontaneous emission of an incoherently pumped ensemble of atoms.
\end{abstract}

\maketitle


\section{\label{sec:introduction} Introduction}
Phase-space descriptions of many-body quantum systems are potentially powerful approaches to study their time evolution. In these approaches, the system's dynamics is mapped from Hilbert space, that scales exponentially with the number of involved degrees of freedom, to the evolution of a quasi-probability function in a quasi-classical phase space of canonical variables whose dimension scales linearly instead. Thereby the computational complexity can be drastically reduced and this class of methods has found a large variety of applications ranging from quantum chemistry, solid-state physics and quantum optics 
\cite{landau_binder_2014, Gull2011,Berne1986,Voth2007, Habershon2013, matsubara_dynamics, Hele2015}. In general these methods rely on a representation of the quantum system in terms of generalized coherent states and an appropriate phase-space distribution function, that can be interpreted as a (quasi) probability function. The temporal evolution of the distribution function is typically governed by a generalized Focker-Planck equation that in turn can be sampled by stochastic trajectories, often resembling classical dynamics. Along this line, stochastic sampling of the generalized positive P-function for bosonic fields \cite{drummond1980generalised} has proven as a powerful method and has been successfully applied to various problems \cite{D_Drummond_2016} of quantum optics \cite{PhysRevLett.58.1841, PhysRevA.41.3930, Drummond_1993, PhysRevA.96.013854, 2018OptCo.427..447O} and Bose-Einstein condensates \cite{PhysRevLett.86.3220, Plimak_2001_Optimization, PhysRevLett.92.040405, PhysRevLett.98.120402, PhysRevE.96.013309, PRXQuantum.2.010319}. Phase-space descriptions of fermionic many-body systems still pose a formidable challenge, but recently Grassmann phase-space methods have been formally introduced and applied \cite{PhysRevA.94.062104, markland}.\\
\indent In this work we introduce a phase-space description based on the positive P representation for bosonic fields interacting with a system of a fixed number of quantum emitters (particles). The proposed formalism will find application in the field of collective light-matter interaction and is extendable to an open quantum system approach, capable of describing decoherence and dissipation due to the interaction with non Markovian baths. The number of particles is supposed to be conserved -- these quantum particles can change their state, but cannot be created or annihilated. Typically, such objects are characterized by transition operators similar to spin \nicefrac{1}{2} $\sigma$ operators in case of two-level atoms. Those transition operators can in turn be represented by bosonic operators by a Jordan-Schwinger transformation \cite{biedenharn1984angular, schwinger1965quantum, klein1991boson, kumar1980theory, carusotto1989dynamics, drobny1992quantum, wu1999schwinger, walls1970quantum} and thus introducing a bozonization of the particle representation. This directly leads to a representation of the system in terms of the positive P-representation for bosons \cite{Olsen2005, Ng_2011}. We extend this strategy to particles represented by more than two levels, with the only limitation inherent to the positive P-representation itself:  the systems needs to be restricted to (at most) two-particle interactions. Practically, the sampling of the resulting generalized Fokker-Planck equations by stochastic trajectories often results in diverging, unbound trajectories, limiting applications to only relatively short propagation times \cite{PhysRevA.55.3014}. This problem can be mitigated by the application of so-called stochastic gauges \cite{2001CoPhC.142..442D, 2002'Deuar_gauge,2003'Drummond_gauge,2006'Deuar_gauge,2005'DeuarPhD}, thereby extending the applicability of the stochastic phase-space sampling method by stochastic trajectories.

The key goal of this work is to establish a map from the traditional quantum-mechanical description to the stochastic formalism. Whereas the quantum-mechanical formulation is based on quantum states and operators, a statistical formalism operates with phase-space distributions, which can be sampled by stochastic processes. 
Our framework opens the way to computationally tractable treatment of collective light-matter interaction involving a large number of many-level quantum emitters. In the hope of our approach being useful in other areas, we provide the equations and their properties without specifying the explicit form of the Hamiltonian.

Our article is organized as follows. In section \ref{sec:quasiprobability distribution}, we introduce a well-known phase-space methodology developed for bosonic fields \cite{drummond1980generalised, drummond2014quantum}. Traditionally, the key objects for the theoretical description of this system are a positive P-function and a corresponding Fokker-Planck equation that determines its temporal evolution and can be sampled with stochastic equations ("trajectories"). Instead, our formalism circumvents the introduction of a Fokker-Planck equation that can be subject to many formal restrictions. For example, Fokker-Planck equation must not contain imaginary values. We directly derive stochastic differential equations that turn out to have an intuitive form. The resulting equations remind canonical equations of classical mechanics and can be directly determined from the system's underlying Hamiltonian. Our prescription to generate the governing stochastic equation of motion holds for a general Hamiltonian. 

In section \ref{sec:II}, we focus on an ensemble of particles (non-relativistic electrons, nuclei, atoms, molecules, quantum emitters, etc. that are considered immovable) and its description in terms of stochastic differential equations. Although the number of particles is assumed to be constant, their respective quantum states can be changed by the interaction with the bosonic field. Based on the Jordan-Swinger transformation \cite{biedenharn1984angular, schwinger1965quantum, klein1991boson, kumar1980theory, carusotto1989dynamics, drobny1992quantum, wu1999schwinger, walls1970quantum}, the particles can be replaced by a system of bosonic fields. This, in turn, allows us to extend the results of section \ref{sec:quasiprobability distribution} and introduce corresponding stochastic equation of motions for the ensemble of particles. Further, we discuss their statistical properties (section \ref{eq: about theta numbers}).

In section \ref{sec: effective density}, we show that the stochastic variables can always be encompassed into objects that we identify as effective density matrices. We derive their equations of motion. The resulting equations are especially convenient when introducing interaction with the reservoir (section \ref{sec: incoherent processes}). If each atom interacts with a separate Markovian reservoir, the equations for the effective density matrices acquire additional drift terms in the style of rate equations. The stochastic terms only originate from the interaction of particle and fields.

In section \ref{sec: stochastic gauges} we discuss the so-called gauge freedom of the stochastic equations of motion:
generally, different distributions can generate the same statistical moments. In \cite{2002'Deuar_gauge,2003'Drummond_gauge,2006'Deuar_gauge,2005'DeuarPhD}, it is shown that there is certain freedom in defining stochastic trajectories, which can be exploited to improve the convergence properties. The underlying technique is called stochastic gauging and is discussed in section \ref{sec: stochastic gauges}. In particular, we show how stochastic gauge transformations can be used to shape the initial conditions possessing inconvenient statistical properties.

In section \ref{sec: discussion}, we discuss future extensions of the formalism. We conclude the article by presenting a series of concrete future applications of our formalism.

\section{\label{sec:quasiprobability distribution}Positive P-representation}


A large-class of quasi-classical treatments of quantum many-body systems relies on the decomposition of the system's density matrix in an appropriate overcomplete basis for bosonic fields \cite{drummond2014quantum}, leading to the introduction of well-known P, R, Q or W functions \cite{PhysRev.131.2766, 1940264, PhysRev.40.749} encoding the evolution of the observables in phase space. The concept of the  R-function was "explicitly" \cite{D_Drummond_2016} extended \cite{Chaturvedi_1977} to a generalized positive P-function \cite{drummond1980generalised} and successfully applied to various problems \cite{D_Drummond_2016, PhysRevLett.58.1841, PhysRevA.41.3930, Drummond_1993, PhysRevA.96.013854, 2018OptCo.427..447O, PhysRevLett.86.3220, Plimak_2001_Optimization, PhysRevLett.92.040405, PhysRevLett.98.120402, PhysRevE.96.013309}. 
If the Hamiltonian involves at most two-particle interaction, the evolution of the corresponding positive P-function can be conveniently sampled with stochastic Ito differential equations\footnote{Under the same condition, W-functions typically require truncating certain terms in the Hamiltonian \cite{Plimak_2001, doi:10.1080/00018730802564254, PhysRevA.80.033624, PhysRevA.96.013854, 2017PhDT........74D}. To avoid it, there are attempts to go beyond stochastic Ito equations \cite{Plimak_2001, Drummond_2014}.}. In this framework, the bosonic field operators are mapped to stochastic complex-valued variables. Whereas the original Hilbert space scales exponentially with respect to the number of bosonic fields, the number of the stochastic variables only grows linearly. The goal of this work is to extend the stochastic phase-space representation beyond bosonic fields.

To recap the scope of stochastic phase-space methods, we start with considering a single bosonic field with arbitrary Hamiltonian
\begin{equation*}
    \hat{H}=\,:\!\!H\!\left(\hat{a}^\dag,\hat{a}\right)\!\!:
\end{equation*}
where colons $:\!...\!:$ denote normal order of the operators. Later, we generalize the resulting equations to the case of several bosonic fields.
Traditional approaches to derive stochastic differential equations sampling the phase space distribution rely on transforming a generalized Fokker-Planck equation into stochastic differential equations. Here, we present a different treatment bypassing the Fokker-Planck equation and we directly introduce stochastic equations of motion for the complex-variable analogues of the bosonic operators.

In a nutshell, the derivation can be outlined in the following procedure. First, we introduce the positive P-distribution (\ref{eq: density matrix decomposition}) that can be interpreted as a probabilistic analogue of the density matrix. We introduce a mapping of operators to complex variables. The resulting P-distribution is a multi-dimensional object in those variables so we sample it by stochastic trajectories, which number scales linearly. To that end, the equations of motion (\ref{eq: the framework of the stochastic equations}) governing the phase-space trajectories are derived, resulting in an intuitive form: the final equations have the form of the canonical equations of classical mechanics (\ref{eq: canonical form}).

Although the resulting equations have been derived for bosonic fields, the same derivation procedure can be extended to describe an ensemble of quantum particles characterized by a many-body wave function or, if in a statistical mixture, by a many-body density matrix. In section \ref{sec:II}, we introduce a bosonization of the many-body wave function. Mapping the ensemble of particles to a system of bosons, the equations derived in this section become applicable to the more general case. 

\subsection{From density matrices to probability distributions}

Consider a bosonic field characterized by creation and annihilation operators $\hat{a}^\dag$ and $\hat{a}$. It is convenient to introduce coherent states or eigenstates of annihilation operator \cite{PhysRev.131.2766}
\begin{equation}
    \label{eq: properties of the coherent states}
    \hat{a}\ket{\alpha}=\alpha\ket{\alpha},
\end{equation}
that compose an overcomplete basis for the bosonic states. Combining ket and bra vectors into projectors \cite{drummond2014quantum}
\begin{equation*}
    \hat{\Lambda}\left(\alpha,\alpha^\dag\right)=\ket{\alpha}\bra{\alpha^{\dag*}}\exp(-\alpha^{\dag}\alpha+|\alpha|^2/2+|\alpha^\dag|^2/2),
\end{equation*}
operators and quantum states can be expanded in the basis of coherent states. Here, $\alpha$ and $\alpha^{\dag}$ are independent complex numbers. The $\Lambda$-projectors are analytical functions with respect to  $\alpha$ and $\alpha^{\dag}$. Thanks to the properties of the coherent states~(\ref{eq: properties of the coherent states}), $\hat{\Lambda}\left(\alpha,\alpha^\dag\right)$ transforms operators into complex numbers
\begin{equation}
        \label{eq: properties of lambda}
        \begin{split}
    \hat{a}\hat{\Lambda}(\alpha&,\alpha^\dag)=\alpha \hat{\Lambda}\left(\alpha,\alpha^\dag\right),\\
    \hat{a}^\dag\hat{\Lambda}\left(\alpha,\alpha^\dag\right)&=\left(\alpha^\dag+\frac{\partial}{\partial\alpha}\right) \hat{\Lambda}\left(\alpha,\alpha^\dag\right),
\end{split}
\end{equation}
which is a key attribute exploited in mapping operators to analogous complex variables. Since the density matrix $\hat{\rho}(t)$ and its evolution completely describe the bosonic field, an optimal way to perform this transition would be to decompose the density matrix into a combination of the $\Lambda$-projectors:
\begin{equation}
    \label{eq: density matrix decomposition}
    \hat{\rho}(t)=\int P(t,\alpha,\alpha^\dag)\hat{\Lambda}(\alpha,\alpha^\dag) d^2\alpha d^2\alpha^\dag.
\end{equation}
The integral in this equation runs over the whole complex planes of $\alpha$ and $\alpha^\dag$\footnote{Choosing integration along particular paths in the complex planes leads to, for example, the notion of the complex P-function often used in analytical investigations \cite{drummond2014quantum}.}. Due to the overcompleteness of the coherent states, there is no unique decomposition of the density matrix (\ref{eq: density matrix decomposition}). However, one can find at least one representation that leads to a positive and real $P(t,\alpha,\alpha^\dag)$ \cite{drummond1980generalised}, which allows interpreting $P(t,\alpha,\alpha^\dag)$ as a probability density function.

This positive, real P-distribution is the key to introducing a probabilistic treatment: in the probabilistic language, expectation values of the operators transform into moments of the probability distribution. This can be directly shown by substituting (\ref{eq: density matrix decomposition}) into the expectation value and making use of the properties of $\hat{\Lambda}\left(\alpha,\alpha^\dag\right)$ (\ref{eq: properties of lambda}). For a general observable $f\!\left(\hat{a}^{\!\dag},\hat{a}\right)$ the expectation value maps to
\begin{equation}
    \label{eq: observables in the stochastic picture}
    \!\!\text{Tr}\!\left(:\!f\!\left(\hat{a}^{\!\dag},\hat{a}\right)\!:\hat\rho\!\left(t\right)\right)=\!\int\!\! f\!\left(\alpha^{\!\dag},\alpha\right) P\!\left(t,\alpha,\alpha^{\!\dag}\right) d^2\alpha^{\!\dag} d^2\alpha.\!\!
\end{equation}
In the probabilistic approach, the complex variables $\alpha^\dag,\alpha$ thus replace the operators $\hat{a}^\dag,\hat{a}$ and  $P\!\left(t,\alpha, \alpha^\dag\right)$ takes over the role of the density matrix $\hat\rho\!\left(t\right)$. Although the transition from operators to complex variables seemingly has simplified the problem, one has to bear in mind that the distribution density is a multi-dimensional object. Generally, the number of the arguments of the distribution density grows linearly with the degrees of freedom of the bosonic fields. As such, the problem to calculate the evolution of $P\!\left(t,\alpha, \alpha^\dag\right)$ in time is still cumbersome. In this work, we opt for a stochastic approach, to sample the time-dependent P-distribution by stochastic processes. In fact, each complex argument $\alpha^\dag,\alpha$ of the bosonic field will describe stochastic trajectories.  Time-dependent expectation values of operators can, in turn, be calculated by averages of the stochastic trajectories (or moments thereof). In our formalism, the P-function is merely an intermediate object that allows mapping of the operator equations of motion onto appropriate equations of motion for the respective complex variables, in their average representing operator expectation values. The hope of the stochastic approach is thus to reduce the numerical complexity of the problem. 

\subsection{Sampling P-distribution by stochastic trajectories}\label{sec:fields}

For a moment, consider a classical single-mode field defined by its positive and negative-frequency mode amplitude $\alpha(t),\alpha^\dag(t)$. For this deterministic, classical field, the probability distribution function is a product of two delta-functions
\begin{equation}
    \label{eq: P decomposition classical}
    P(t,\alpha,\alpha^\dag)=\delta(\alpha(t)-\alpha)\delta(\alpha^\dag(t)-\alpha^\dag).\footnote{Note, that the variables without the time dependency $\alpha,\alpha^\dag$ are the arguments of the distribution function. Further note that the arguments of the delta-functions are complex, $\delta(z)$ stands for a product of two ordinary delta-functions for real $\delta(z')$ and imaginary $\delta(z'')$ components of the argument.}
\end{equation}
For a certain class of "quasi-classical" quantum fields, the distribution function may look like smeared delta-functions that are spread over a some small, finite region in the phase space. Evidently, for these distribution functions of small support, we only need to evaluate it in small, compact area in phase-space. The distribution can be seen as a combination of some independent trajectories, starting from a compact area of initial conditions. For "quasi-classical" systems, these trajectories have strong similarities with classical mechanics and are the key to developing numerically tractable computational methods.

For a general bosonic field, the support of the P-function might not be centered around classical trajectories. The hope is, however, that launching stochastic trajectories from an appropriately sampled area of initial conditions, leads to a well-sampled density function at later times. 

To mimic the smeared-out, diffusing distribution function, we replace classical variables $\alpha(t),\alpha^\dag(t)$ in equation (\ref{eq: P decomposition classical}) with stochastic Ito processes, enhancing the complexity of the trajectories to also describe quantum effects. In general, their equations of motion include not only deterministic contributions $A(t)$ and $A^\dag(t)$, but also complex Gaussian white noise terms $\zeta\!\left(t\right)$ and $\zeta^\dag\!\left(t\right)$
\begin{align}
    \label{eq: the framework of the stochastic equations}
    \begin{split}
    \frac{d\alpha\!\left(t\right)}{dt}\!=&A(t)+\zeta\!\left(t\right),\\
        \frac{d\alpha^\dag\!\left(t\right)}{dt}=&A^\dag(t)+\zeta^\dag\!\left(t\right).
    \end{split}
\end{align}
Since we consider the noise terms in the sense of Ito, their first moments vanish. The functional form  of the deterministic terms $A(t)$, $A^\dag(t)$ and the correlation properties of the noise terms $\zeta\!\left(t\right)$, $\zeta^\dag\!\left(t\right)$ are solely determined by the physical properties of the system and its Hamiltonian and will be derived in the next subsection. Noise terms make every realization of stochastic equations unique.
Within the stochastic approach, the P-function at each time point $t$ can be recovered by an ensemble average over stochastic trajectories. Each point in phase space is represented with a product of delta functions, "counting" the number of trajectories that pass through that point in phase-space at time $t$:   
\begin{equation}
    \label{eq: P decomposition}
    P(t,\alpha,\alpha^\dag)=\Big\langle\delta(\alpha(t)-\alpha)\delta(\alpha^\dag(t)-\alpha^\dag)\Big\rangle.
\end{equation}
Being sampled by the stochastic variables, the equation of motion for $P(t,\alpha,\alpha^\dag)$ is a type of Fokker-Planck equation -- the usual equation for a probabilistic treatment of bosonic systems \cite{drummond2014quantum, scully1999quantum}. 

Sampling $P(t,\alpha,\alpha^\dag)$ with a set of stochastic trajectories $\alpha(t)$ and $\alpha^\dag(t)$ and applying the properties of the delta-functions, we complete the transition from the quantum-mechanical formulation to the stochastic one by the calculation of expectation values of a general observable:
\begin{eqnarray}
    \label{eq: the transition}
    \!\!\text{Tr}\!\left\{:\!f\!\left(\hat{a}^{\!\dag},\hat{a}\right)\!:\hat\rho\!\left(t\right)\right\}&=&\!\int\!\! f\!\left(\alpha^{\!\dag},\alpha\right) P\!\left(t,\alpha,\alpha^{\!\dag}\right) d^2\alpha^{\!\dag} d^2\alpha\!\!\nonumber\\
    &=&\Big\langle f\!\left(\alpha^{\!\dag}(t),\alpha(t)\right)\!\Big\rangle.
\end{eqnarray}
The bracket $\Big\langle .... \Big\rangle$ denotes the ensemble average over stochastic trajectories.

Within our approach, the density matrix $ \hat{\rho}(t)$ can be directly sampled by the stochastic trajectories, without explicitly constructing $P(t,\alpha,\alpha^\dag)$. Using the ansatz for $P(t,\alpha,\alpha^\dag)$ (\ref{eq: P decomposition}) and the decomposition for the density matrix (\ref{eq: density matrix decomposition}), we arrive at
\begin{equation}
    \label{eq: density matrix through the trajectories}
    \hat{\rho}(t)=\Big\langle\hat{\Lambda}\!\left(\alpha(t),\alpha^\dag(t)\right)\Big\rangle.
\end{equation}
This expression is especially useful (see next section), when deriving the equations of motion for the stochastic variables. Note that by neglecting the stochastic components, assuming classical trajectories and omitting the ensemble average, the density matrix is sampled by a single $\Lambda$-projector parametrized with time. The statistical sampling is a convenient way to construct a superposition of the $\Lambda$-projector. In the next subsection, we show that this sampling is exact for a general many-body Hamiltonian with two particle interactions.

Note, that we have not provided any proof that the stochastic interpretation of the density matrix (\ref{eq: density matrix through the trajectories}) is always possible. In principle, the underlying distribution can stretch to such extend that any meaningful Monte Carlo sampling is impossible, which is similar to the boundary problem in the traditional formulation based on the Fokker-Planck equation \cite{PhysRevA.55.3014}. 

\subsection{Stochastic equations of motion}
Let us derive the equations that govern the evolution of the stochastic variables. We start from the Liouville - von Neumann equation of motion (master equation) that determines the temporal evolution of the density matrix:
\begin{equation}
\label{eq: initial master equation}
    \frac{\partial \hat{\rho}(t)}{\partial t}=-\frac{i}{\hbar}\left[\hat{H},\hat{\rho}(t)\right].
\end{equation}
Relation (\ref{eq: density matrix through the trajectories}) links the density matrix to the ensemble average over stochastic trajectories. Inserting it in the master equation and using the properties of the $\Lambda$-projectors (\ref{eq: properties of lambda}) leads to the following equation
\begin{equation}
    \label{eq: equation for the averaged lambda}
    \frac{\partial}{\partial t}\Big\langle\hat{\Lambda}\!\left(\alpha(t),\alpha^\dag(t)\right)\Big\rangle=\frac{i}{\hbar}\Big\langle \mathcal{L}\,\hat{\Lambda}\!\left(\alpha(t),\alpha^\dag(t)\right)\Big\rangle,
\end{equation}
For convenience, we introduce an operator $\mathcal{L}$ defined by
\begin{multline}
    \label{eq: decomposition of L}
    \mathcal{L}=\frac{\partial H}{\partial \alpha}\frac{\partial }{\partial\alpha^\dag}-\frac{\partial H}{\partial \alpha^\dag}\frac{\partial }{\partial\alpha}\\+\frac{1}{2}\frac{\partial^2 H}{\partial \alpha^2}\frac{\partial^2}{\partial\alpha^{\dag2}}-\frac{1}{2}\frac{\partial^2 H}{\partial \alpha^{\dag 2}}\frac{\partial^2}{\partial\alpha^2}+...,
\end{multline}
where $H$ is the stochastic analog of the Hamiltonian operator:
\begin{equation}
    \label{eq: stochastic Hamiltonian}
    \hat{H}=\,:\!\!H\!\left(\hat{a}^\dag,\hat{a}\right)\!\!: \quad \to \quad
    H(t)=H\!\left(\alpha^\dag(t),\alpha(t)\right).
\end{equation}
The detailed derivation of equation\ (\ref{eq: equation for the averaged lambda}) and the general expression of $\mathcal{L}$ is provided in Appendix \ref{app: derivation of L}. In equation\ (\ref{eq: decomposition of L}) we have listed terms up to the second derivative in $\alpha,\alpha^\dag$. Note, that $\mathcal{L}$ does not contain mixed derivatives with respect to the variables $\alpha,\alpha^\dag$. The terms with the first-order derivatives compose a convective derivative, characterising the flow of the distribution density. The second-order derivatives have a diffusive effect. 

Now, let us have a look at equation (\ref{eq: equation for the averaged lambda}) from a different perspective: We derived it, starting from the master equation. Alternatively, we can directly evaluate the total temporal derivative of the left-hand side of (\ref{eq: equation for the averaged lambda}). As previously, let us first assume that the trajectories have no stochastic component, so that equations (\ref{eq: the framework of the stochastic equations}) contain only deterministic terms $\frac{d\alpha}{dt} =:A(t)$ and $\frac{d\alpha^\dag}{dt} =:A^\dag(t)$. Then, the right-hand side of the equation (\ref{eq: equation for the averaged lambda}) contains only the first-order derivatives
\begin{multline}
\label{eq: classical contribution}
\frac{\partial}{\partial t}\Big\langle\hat{\Lambda}\!\left(\alpha(t),\alpha^\dag(t)\right)\Big\rangle\\=\Big\langle \left[A^\dag(t)\frac{\partial}{\partial \alpha^\dag}+A(t)\frac{\partial}{\partial \alpha}\right]\,\hat{\Lambda}\!\left(\alpha(t),\alpha^\dag(t)\right)\Big\rangle.
\end{multline}
Using the classical trajectories, we can reproduce only the first two terms in the decomposition of $\mathcal{L}$ (\ref{eq: decomposition of L})
\begin{subequations}
\begin{equation}
\label{eq: canonical form}
\begin{gathered}
     \left(\frac{d\alpha\!\left(t\right)}{dt}\right)_{\text{det.}}\!\!\!\!=A(t)=-\frac{i}{\hbar}\frac{\partial H}{\partial \alpha^\dag},\\ \left(\frac{d\alpha^\dag\!\left(t\right)}{dt}\right)_{\!\text{det.}}\!\!\!\!\!=A^\dag(t)=\frac{i}{\hbar}\frac{\partial H}{\partial \alpha}.
\end{gathered}
\end{equation}
These expressions remind Hamilton's equations for canonical variables. The deterministic terms of the equation of motion of the complex variables thus follow classical mechanics. By converse argument, the inclusion of the noise terms must recover the next-to leading terms in the decomposition of $\mathcal{L}$ (\ref{eq: decomposition of L}) responsible for the quantum properties. Indeed, according to a well-known Ito's lemma \cite{gardiner2004handbook}, the stochastic processes have unusual differentiation rules. Consequently, the expression (\ref{eq: classical contribution}) is accompanied with the terms dependent on the correlation properties of the noise terms $\zeta\!\left(t\right)$ and $\zeta^\dag\!\left(t\right)$. The detailed derivation of the stochastic equations of motion can be found in Appendix \ref{app: ito}. Choosing the following correlation properties of the noise terms
\begin{equation}
\label{eq: correlation properties for two bosons}
\begin{split}
    \langle\zeta\!\left(t\right)\zeta\!\left(t'\right)\rangle&=-\frac{i}{\hbar}\frac{\partial^2 H}{\partial \alpha^{\dag 2}}\delta(t-t'),\\
    \langle\zeta^\dag\!\!\left(t\right)\zeta^\dag\!\!\left(t'\right)\rangle&=\frac{i}{\hbar}\frac{\partial^2 H}{\partial \alpha^2}\delta(t-t'),\\
    \langle\zeta^\dag\!\left(t\right)\zeta\!\left(t'\right)\rangle&=0,
\end{split}
\end{equation}
\end{subequations}
we can recover the diffusive second-order derivatives in equations (\ref{eq: equation for the averaged lambda}), (\ref{eq: decomposition of L}). Since the noise terms are Gaussian, we need only first and second momenta to define their statistics. The quantum stochastic equations of motion are therefore given by
\begin{align}
\label{eq: full bosonic equations one mode}
    \begin{split}
    \frac{d\alpha\!\left(t\right)}{dt}\!=\!-\frac{i}{\hbar}\frac{\partial H}{\partial \alpha^\dag}+\zeta\!&\left(t\right),\\
        \frac{d\alpha^\dag\!\left(t\right)}{dt}=\frac{i}{\hbar}\frac{\partial H}{\partial \alpha}+\zeta^\dag\!&\left(t\right).
    \end{split}
\end{align}

Unfortunately, only the systems governed by Hamiltonians with a polynomial functional dependence of creation- and annihilation operators up to at most order 2 can be solved exactly by the stochastic-trajectory approach. Despite this limitation, this still covers a large class of many-body Hamiltonians. For systems with Hamiltonians of higher polynomial degree in the annihilation/creation operators, the stochastic approach is an approximate solution. For those systems, one effectively drops the terms beyond second-order derivatives in equation (\ref{eq: decomposition of L}), which might also be justifiable in many cases.

\subsection{Multi-mode bosonic fields}
Let us now extend the derived expressions to the case of an arbitrary number of modes of the boson field, with mutually independent pairs of bosonic creation and annihilation operators $\hat{a}_i^\dag$, $\hat{a}_i$ and a general Hamiltonian $\hat{H}$.
By analogy with the single-mode case, we introduce corresponding stochastic variables $\alpha_i^\dag$, $\alpha_i$ and the stochastic analog of the Hamiltonian
\begin{subequations}
    \label{eq: Hamiltonian for the many field variables}
\begin{equation}
\begin{gathered}
    \hat{H}=\,:\!\!H\!\left(...,\hat{a}^\dag_i,\hat{a}_i,...\right)\!\!: \\ \downarrow \\
    H(t)=\,H\!\left(...,\alpha^\dag_i(t),\alpha_i(t),...\right)\,
\end{gathered}  
\end{equation}
The stochastic equations of motion can be easily obtained by adding corresponding indices to the previously derived equations (\ref{eq: full bosonic equations one mode})
\begin{align}\label{eq: equations of motion for the many field variables}
    \begin{split}
    \frac{d\alpha_i\!\left(t\right)}{dt}\!=\!-\frac{i}{\hbar}\frac{\partial H}{\partial \alpha^\dag_i}+\zeta_i\!&\left(t\right),\\
        \frac{d\alpha^\dag_i\!\left(t\right)}{dt}=\frac{i}{\hbar}\frac{\partial H}{\partial \alpha_i}+\zeta^\dag_i\!&\left(t\right).
    \end{split}
\end{align}
For correlation properties of the noise (\ref{eq: correlation properties for two bosons}) we get:
\begin{equation}
\begin{split}
    \label{eq: correlation properties for the noise variables}
    \langle\zeta_i\!\left(t\right)\zeta_j\!\left(t'\right)\rangle&=-\frac{i}{\hbar}\frac{\partial^2 H}{\partial \alpha^\dag_i\partial \alpha^\dag_j}\delta(t-t'),\\
    \langle\zeta^\dag_i\!\left(t\right)\zeta^\dag_j\!\left(t'\right)\rangle&=\frac{i}{\hbar}\frac{\partial^2 H}{\partial \alpha_i\partial \alpha_j}\delta(t-t'),\\
    \langle\zeta^\dag_i\!\left(t\right)\zeta_j\!\left(t'\right)\rangle&=0.
\end{split}
\end{equation}
We thus have derived a prescription to generate the stochastic canonical equations of motion (\ref{eq: equations of motion for the many field variables}) for a given arbitrary bosonic Hamiltonian represented by creation- and annihilation operators. Analogous to equation (\ref{eq: the transition}) the expectation value of a general observable $f$ reads
\begin{multline}
    \!\!\text{Tr}\!\left\{:\!f\!\left(...,\hat{a}^\dag_i,\hat{a}_i,...\right)\!:\hat\rho\!\left(t\right)\right\}\\=\Big\langle f\!\left(...,\alpha^\dag_i(t),\alpha_i(t),...\right) \!\Big\rangle.
\end{multline}
\end{subequations}
Although our formalism has been derived for bosonic fields, it can readily be extended to an ensemble of quantum particles after introduction of an appropriate bosonization. The next section introduces bosonization for an ensemble of particles that directly leads towards the stochastic equations of motion for the corresponding complex variables.
\section{\label{sec:II} Particles}
In the last section, we introduced the stochastic formalism that transformed the master equation of multi-mode bosonic fields into a set of stochastic equations of motion for the complex variable equivalents of the bosonic creation and annihilation operators. Here, we extend this idea to a system of quantum particles that can only
perform transitions from one state to another. Transitions between state $q$ to $p$ can be represented by an extension of the Pauli matrices $\sigma_{pq}\equiv\ket{p}\bra{q}$ \cite{Benedict2018}.

Several stochastic methods to treat fermionic or bosonic particles, represented by a wave functions, have been introduced in the past. Parametrizing the density matrix by fermion coherent states leads to Grassmann phase space methods for fermions \cite{DALTON201612, DALTON2017268, Polyakov2016, PhysRevA.64.063409, Kidwani_2020}. Gaussian phase-space representations \cite{PhysRevLett.93.260401, PhysRevA.68.063822, PhysRevB.73.125112} provide a unified method for bosons and fermions alike. Transition $\sigma$-operators allow for the construction of coherent states, naturally extending the concept of the positive P-representation to atoms or particles \cite{PhysRevA.78.052108}. Notably, the latter method has been applied to treat interacting spin systems \cite{PhysRevB.88.144304} and spin-boson networks \cite{Mandt_2015}. Another way of characterizing atoms is to introduce collective transition operators and a corresponding characteristic function \cite{PhysRevA.23.2563, PhysRevA.34.3166, PhysRevA.35.3832, PhysRevA.40.5135, drummond1991quantum, PhysRevA.77.012323, Andreasen:09, PhysRevA.85.013835}, a method that is formally exact in the limit of an infinite number of atoms. Typically these methods only consider two-level atoms\footnote{Ref. \cite{Andreasen:09, PhysRevA.85.013835} are rare examples with three- and four-level atoms respectively.} and at most for a one-dimensional geometry. The latter method for embedding atoms \cite{PhysRevA.23.2563, PhysRevA.34.3166, PhysRevA.35.3832, PhysRevA.40.5135, drummond1991quantum, PhysRevA.77.012323, Andreasen:09, PhysRevA.85.013835} is based on a coarse-grained model which is not always justifiable\footnote{Although there is a similar approach \cite{PhysRevA.38.4073} suitable for any number of two-level atoms, it requires additional auxiliary variable or third-order noise terms.}. Yet another framework to the stochastic description of atoms \cite{Olsen2005, Ng_2011, PhysRevA.102.012219, Huber_2021} is based on the Jordan-Schwinger transformation \cite{biedenharn1984angular, schwinger1965quantum, klein1991boson, kumar1980theory, carusotto1989dynamics, drobny1992quantum, wu1999schwinger, walls1970quantum}. Mathematically, this corresponds to introducing boson operators under the second quantization formalism, which we will adopt and is described in this section.

First, we consider only the simple case of single atom to be generalized to an arbitrary number of particles. In section \ref{subsec: Hamiltonian}, we introduce the Hamiltonian $\hat{H}$ of an ensemble of particles interacting with bosonic fields, discuss its structure and apply second quantization to it. Having formulated the system particles$+$fields in terms of a bosonic representation, we adopt the formalism derived in the previous section to derive stochastic trajectory equations of motion. Restricting the interaction between particles and fields to two-particle interaction, the resulting equations are in principle exact.

By the nature of the $\sigma$-operators, the particle number is conserved. This results in interesting properties of the bosonic description. First, we identify a certain freedom in decomposing the operators for particles in terms of bosonic operators (\ref{eq: freedom for the bosons}). This leads to stochastic trajectories possessing non-trivial statistical properties (\ref{eq: theta-numbers}). Based on the example of a system of initially non-interacting atoms, we show explicitly how this non-trivial statistics is reflected in the trajectories  (Appendix \ref{app: uncorrelated atoms}). Another important property is that the bosonic operators associated with particles appear always in pairs. In section \ref{sec: effective density}, this property is used to introduce so-called effective density-matrix elements. 
\subsection{$\sigma$-projectors}

Let us start with the simple example of one particle that can be characterized by a state vector $\ket{\Psi}$. We expand this state in terms of an orthonormal basis $\{\ket{p}\}$
\begin{subequations}
\label{eq: sigma-projectors}
\begin{equation}
\label{eq: state decomposition}
    \ket{\Psi}=\sum_p C_p\ket{p}.
\end{equation}
Any operator $\hat{f}$ acting on the particle can be expressed through the projectors $\hat{\sigma}_{pq}\equiv\ket{p}\bra{q}$ by
\begin{equation}
    \label{eq: operator decomposition}
    \hat{f}=\sum_{p,q}f_{pq}\ket{p}\bra{q}=\sum_{p,q}f_{pq}\hat{\sigma}_{pq}.
\end{equation}
\end{subequations}
The $\sigma$-projectors have a non-trivial algebra with commutation relations
\begin{equation}\label{eq:sigma commutators}
    \left[\hat{\sigma}_{pq},\hat{\sigma}_{rs}\right]=\hat{\sigma}_{ps}\delta_{rq}-\hat{\sigma}_{rq}\delta_{ps}.
\end{equation}
Evidently, the commutator is more complex as compared to simple bosonic commutation relations. This hinders a direct extension of the results  of section \ref{sec:fields} to the case of an ensemble of indistinguishable particles. Despite this complexity, the $\sigma$-operators have been previously used for developing stochastic phase-space sampling techniques of a many-particle ensemble \cite{drummond1991quantum, Hetet2008}. In these approaches, the non-trivial commutator generated additional terms in the equation of motion for the positive P distribution (even for particles only underlying to two-particle interactions) that in turn had to be neglected to get the form of Fokker-Planck equation, thereby limiting the applicability and validity of the method.
Another challenge to represent the quantum system by $\sigma$-operators becomes evident for multi-level particles -- the ordering of operators, important for the interpretation of statistical averaging. The commutation relations (\ref{eq:sigma commutators}) do not suggest an evident ordering prescription, and can therefore be individually defined for a particular application (see, for example, \cite{Hetet2008}). 
We aim for a generally applicable stochastic approach, without any ad hoc approximations, that relies on the introduction of new operators.
\subsection{$c$-operators}\label{sec: c-operators}
To develop a unified approach, treating particles and fields alike, we simplify the structure of the introduced $\sigma$-operators. To that end we introduce $c$-operators, in the spirit of second quantization. As mentioned earlier, the action of $\sigma$-operators can be seen as an annihilation of the atom in a particular state with a subsequent creation of the atom in another state. Consequently, we introduce an intermediate state $\ket{\o}$ and creation operators ${\hat{c}^\dag_p}$ turning this state into the basis state~${\ket{p}}$
\begin{subequations}
\label{eq: a new interpretation}
\begin{equation}
    \ket{\Psi}=\sum_p C_p^{(0)}\ket{p}=\sum_p C_p^{(0)}\hat{c}^\dag_p\ket{\o}.
\end{equation}
Treating $\ket{\o}$ as a vacuum state and assuming bosonic or fermionic commutation relation $[\hat{c}_p,\hat{c}^\dag_q]_{\pm}=\delta_{pq}$, one can mimic the action of $\sigma$-operators with a pair of creation and annihilation operators $\hat{\sigma}_{pq}=\hat{c}_p^\dag\hat{c}_q$
\begin{multline}
    \hat{\sigma}_{pq}\ket{r}=\hat{c}_p^\dag\hat{c}_q\hat{c}^\dag_r\ket{\o}=\hat{c}_p^\dag[\hat{c}_q,\hat{c}^\dag_r]_{\pm}\ket{\o}\\=\hat{c}_p^\dag\delta_{qr}\ket{\o}=\ket{p}\delta_{qr}.
\end{multline}
\end{subequations}
In contrast to $\sigma$-projectors, $c$-operators span a large auxiliary space with an occupation-number basis: Every $c$-operator has a corresponding occupation number that it can modify. Initial basis states $\ket{p}$ only correspond to a relatively small subspace of states with a total occupation number equal to one:
\begin{center}
\begin{tabular}{ c c c }
 single state &  & occupation number \\  
  representation &  & representation \\  
     &  &  \\ 
   $\ket{p}$ &  & $\ket{0,\dots,n_p\!\!=\!1,\dots,0}$ \\ 
   &  &  \\  
 $\ket{1}$ &  & $\ket{\,1,\,0\,,\,0\,,\,0\,,\,\dots,\,0\,}$ \\ 
 $\ket{2}$ & $\longrightarrow$ & $\ket{\,0,\,1\,,\,0\,,\,0\,,\,\dots,\,0\,}$  \\  
 $\ket{3}$ &  & $\ket{\,0,\,0\,,\,1\,,\,0\,,\,\dots,\,0\,}$  \\ 
 $\ket{4}$ &  & $\ket{\,0,\,0\,,\,0\,,\,1\,,\,\dots,\,0\,}$  \\ 
 \dots &  & \dots  \\
\end{tabular}
\end{center}
In this work we focus on bosonic commutation relations for the $c$-operators, which results in a direct applicability of the results of section \ref{sec:fields}. The transformation of the operator basis expansion from $\sigma$- to $c$-operators is similar to a Jordan–Schwinger mapping \cite{biedenharn1984angular, schwinger1965quantum, klein1991boson, kumar1980theory, carusotto1989dynamics, drobny1992quantum, wu1999schwinger, walls1970quantum}, thereby avoiding the non-trivial commutation relations of the $\sigma$-projectors and providing a unique operator ordering. Identical results can be also obtained by directly applying the traditional second quantization formalism \cite{landau1977} developed for an ensemble of identical quantum particles \cite{gegg2016efficient, gegg2017identical, gegg2017psiquasp, shammah2018open, bolanos2015algebraic, chase2008collective, xu2013simulating, hartmann2016generalized}.  An alternative approach to bosonization can be found in \cite{https://doi.org/10.48550/arxiv.2207.14234}. Instead of considering the Hilbert space of pure states, the occupation numbers \cite{gegg2016efficient, gegg2017identical, gegg2017psiquasp} and boson operators \cite{bolanos2015algebraic} can be introduced at the level of Liouville space. Such a picture allows for a simpler and intuitive treatment of open systems, at the cost of higher dimensionality of Liouville space and a resulting increase in the number of boson operators. 
\subsection{A system of particles}
Here we extend our considerations to a system consisting of many quantum particles. Each particle, numerated by the indices $a,b...$ obtains a set of $\sigma$-projectors and corresponding independent boson $c$-operators:
\begin{equation}
    \label{eq: sigma operator}
    \hat{\sigma}_{a,pq}=\hat{c}_{a,p}^\dag\hat{c}_{a,q}, \quad  \left[\hat{c}_{a,p},\hat{c}_{b,q}^\dag\right]=\delta_{ab}\delta_{pq}.
\end{equation}
 Generalizing equation\ (\ref{eq: operator decomposition}), a single-particle operator $\hat{f}_a$ acting on particle $a$ is represented by
\begin{subequations}
\label{eq: general decomposition of an operator}
\begin{equation}
    \label{eq: operator expressed through the sigma operators}
    \begin{split}
    \hat{f}_a=\sum_{p,q}f_{a,pq}\hat{\sigma}_{a,pq}=\sum_{p,q}f_{a,pq}\hat{c}_{a,p}^\dag\hat{c}_{a,q},\\
    f_{a,pq}=\bra{p}_a\hat{f}_a\ket{q}_a.
    \end{split}
\end{equation}
A general operator $\hat{F}$ is expressed as
\begin{multline}
 \label{eq: general decomposition of an operator (sigma)}
    \hat{F}=F^{(0)}+\sum_{a,p,q}F_{a,pq}^{(1)}\hat{\sigma}_{a,pq}+\\+\frac{1}{2}{\sum_{\substack{a,p,q \\a',p',q'}}}^{\!\!\!\bm{'}}F_{a,pq;a'p'q'}^{(2)}\hat{\sigma}_{a,pq}\hat{\sigma}_{a',p'q'}+\dots,
\end{multline}
Here, we avoid self-interaction of the particles assuming $a\neq a'$, which is indicated by the prime on the sum. In equation (\ref{eq: general decomposition of an operator (sigma)}), the decomposition does not include the products of $\sigma$-operators corresponding to the same atoms. For the sake of consistency, we call representation (\ref{eq: general decomposition of an operator (sigma)}) as \textit{normal ordering} for $\sigma$-operators, because it directly leads to a normal ordered expression in terms of $c$-operators:
\begin{multline}
    \label{eq: general decomposition of an operator (c)}
    \hat{F}=F^{(0)}+\sum_{a,p,q}F_{a,pq}^{(1)}\hat{c}^\dag_{a,p}\hat{c}_{a,q}+\\+\frac{1}{2}{\sum_{\substack{a,p,q \\a',p',q'}}}^{\!\!\!\bm{'}}F_{a,pq;a'p'q'}^{(2)}\hat{c}^\dag_{a',p'}\hat{c}^\dag_{a,p}\hat{c}_{a,q}\hat{c}_{a',q'}+\dots.
\end{multline}
\end{subequations}
If the operator $\hat{F}$ is not in normal order in terms of $\sigma$-projectors, the decomposition through $c$-operators may contain two consecutive annihilation operators $\hat{c}_{a,p}\hat{c}_{a,q}$ acting on particle $a$. This double action of annihilation operators on a one-particle state results in zero. We therefore drop the normal ordered terms including such products, effectively introducing a fermion-like property of the $c$-operators
\begin{equation}
\label{eq: freedom for the bosons}
    \hat{c}_{a,p}\hat{c}_{a,q}\to 0
\end{equation}
As a result, every operator in Hilbert space of fixed particle number can be represented by the expansion (\ref{eq: general decomposition of an operator (c)}). This fermion-like property results in non-trivial statistics of the associated stochastic trajectories, as will be derived in the following.

\subsection{Hamiltonian of particles and fields}
\label{subsec: Hamiltonian}

The key object characterizing the evolution of the system is the Hamiltonian. We consider an arbitrary Hamiltonian $\hat{H}$ of the composed system of boson fields (photons, phonons, etc.) and the ensemble of (potentially also interacting) particles. The bosonic fields are evidendtly represented by bosonic creation and annihilation operators $\hat{a}_i^\dag$, $\hat{a}_i$. Operators acting on particles are first expanded in terms of $\sigma$-operators. Subsequently, the $\sigma$-operators are decomposed in terms of $c$-operators according to equation\ (\ref{eq: sigma operator}), to arrive at the fully bosonised expression of the particle subsystem. Furthermore, in the view of the stochastic approach, the Hamiltonian $\hat{H}$ is normally ordered (indicated with colons $:\!...\!:$) and we arrive at
\begin{equation}
    \label{eq: Hamiltonian}
    \hat{H}=\,:\!\!H\!\left(...,\hat{a}^\dag_i,\hat{a}_i,...,\hat{c}^\dag_{a,p},\hat{c}_{a,p},...\right)\!\!:
\end{equation}
Additionally, we omit all terms containing a product of two consecutive annihilation operators, see equation\ (\ref{eq: freedom for the bosons}). The Hamiltonian, thus, has the general bosonic form, allowing direct mapping onto the stochastic formalism according to equations\ (\ref{eq: Hamiltonian for the many field variables}).
As discussed earlier, the stochastic approach describes the exact evolution for Hamiltonians with at most two creation or two annihilation operators pertaining to the same field-mode or particle. Two-particle interactions and terms that in the realm of photon-matter interaction come into play via minimal coupling are therefore treated exactly:
\begin{equation*}
    \begin{split}
            \hat{\sigma}_{a,pq}\hat{\sigma}_{b,rs}, \quad\quad\hat{\sigma}_{a,pq}\hat{a}_i, \quad&\quad \hat{\sigma}_{a,pq}\hat{a}_i^\dag, \quad\quad \hat{\sigma}_{a,pq}\hat{a}_i^\dag\hat{a}_j,\\
    \big\downarrow&\\
    \hat{c}_{a,p}^\dag\hat{c}_{a,q}\hat{c}_{b,r}^\dag\hat{c}_{b,s}, \,\,\,\hat{c}_{a,p}^\dag\hat{c}_{a,q}\hat{a}_i, \,\,\,&\,\,\, \hat{c}_{a,p}^\dag\hat{c}_{a,q}\hat{a}_i^\dag, \quad \hat{c}_{a,p}^\dag\hat{c}_{a,q}\hat{a}_i^\dag\hat{a}_j.
    \end{split}
\end{equation*}
Here, the first class of terms corresponds to two-particle interactions, the second (third) class to absorption (emission) of a boson by interaction with a particle. The last class describes scattering of the bosonic field by a particle. Interparticle and particle-field interactions being described beyond these terms, would result in an only approximate evolution by stochastic trajectories. The reader can find particular examples of bosonization of atoms and fermions in Appendix~\ref{app: examples of the bosonization}.
\subsection{Stochastic equations for particles}
At this point, we have all the needed ingredients: the total system's Hamiltonian expressed in boson operators (\ref{eq: Hamiltonian}) and the stochastic formalism suitable for such systems (\ref{eq: equations of motion for the many field variables})-(\ref{eq: correlation properties for the noise variables}). By analogy with the bosonic field, stochastic complex variables $C_{a,p}(t)$ and $C_{a,p}^\dag(t)$ are introduced to represent the operators $\hat{c}_{a,p}(t)$ and $\hat{c}_{a,p}^\dag(t)$, and their stochastic equations of motion are analogous to equations\ (\ref{eq: equations of motion for the many field variables}). To that end we introduce a stochastic Hamiltonian in accordance with equation (\ref{eq: Hamiltonian})
\begin{subequations}
\label{eq: stochastic equations for the particles}
\begin{equation}
\label{eq: total stochastic hamiltonian}
   \!\! H(t)=H\!\left(...,\alpha^\dag_i(t),\alpha_i(t),...,C^\dag_{a,p}(t),C_{a,p}(t),...\right).\!\!
\end{equation}
The stochastic equations of motion for the particles, in analogy to equations\ (\ref{eq: equations of motion for the many field variables}) thus read
\begin{align}
\label{eq: equations for C-variables}
\begin{split}
        \frac{dC_{a,p}\!\left(t\right)}{dt}&=-\frac{i}{\hbar}\frac{\partial H}{\partial C_{a,p}^\dag}+\xi_{a,p}\!\left(t\right),\\
        \frac{dC^\dag_{a,p}\!\left(t\right)}{dt}&=\frac{i}{\hbar}\frac{\partial H}{\partial C_{a,p}}+\xi_{a,p}^\dag\!\left(t\right)
\end{split}
\end{align}
In addition to the deterministic term represented by the derivatives of the Hamiltonian, noise terms $\xi_{a,p}\!\left(t\right)$, $\xi_{a,p}^\dag\!\left(t\right)$ have to be introduced. By analogy with equation\ (\ref{eq: correlation properties for the noise variables}), the noise is characterized by the correlation functions
\begin{equation}
    \label{eq: correlation properties of the noise terms f}
\begin{split}
    \langle\xi_{a,p}\!\left(t\right)\xi_{a',p'}\!\left(t'\right)\rangle&=-\frac{i}{\hbar}\frac{\partial^2 H}{\partial C_{a,p}^\dag\partial C_{a',p'}^\dag}\delta(t-t'),\\
    \langle\zeta_{i}\!\left(t\right)\xi_{a',p'}\!\left(t'\right)\rangle&=-\frac{i}{\hbar}\frac{\partial^2 H}{\partial \alpha_{i}^\dag\partial C_{a',p'}^\dag}\delta(t-t'),\\
    \langle\xi^\dag_{a,p}\!\left(t\right)\xi^\dag_{a',p'}\!\left(t'\right)\rangle&=\frac{i}{\hbar}\frac{\partial^2 H}{\partial C_{a,p}\partial C_{a',p'}}\delta(t-t'),\\
    \langle\zeta^\dag_{i}\!\left(t\right)\xi^\dag_{a',p'}\!\left(t'\right)\rangle&=\frac{i}{\hbar}\frac{\partial^2 H}{\partial \alpha_{i}\partial C_{a',p'}}\delta(t-t').
\end{split}
\end{equation}
All other second-degree correlators vanish:
\begin{multline}
\label{eq: correlation properties of the noise terms f (ad)}
        \langle\xi_{a,p}\!\left(t\right)\xi^\dag_{a',p'}\!\left(t'\right)\rangle=\langle\zeta_{i}\!\left(t\right)\xi^\dag_{a',p'}\!\left(t'\right)\rangle\\=\langle\zeta_{i}^\dag\!\left(t\right)\xi_{a',p'}\!\left(t'\right)\rangle=0.
\end{multline}
\end{subequations}
Although, at first sight, the resulting equations do not seem to differ substantially from the equations\ (\ref{eq: equations of motion for the many field variables}) for the bosonic fields, they posses two important distinctive features with drastic consequences.
First, due to the origin of the $c$-operators linked to the $\sigma$-projectors, the annihilation operators $\hat{c}_{a,i}$ and $\hat{c}_{a,i}^\dag$ pertaining to one particle and hence $C_{a,p}(t)$ and $C_{a,q}^\dag(t)$) appear always pairwise. This implies, that the products $C_{a,p}(t)C_{a,q}^\dag(t)$ can be represented by new variables  
\begin{equation*}
  \rho_{a,pq}(t)\equiv C_{a,p}(t)C_{a,q}^\dag(t)
\end{equation*}
that can be interpreted as an effective density matrix pertaining to particle $a$ (section \ref{sec: effective density}). Second, the $c$-operators have the fermion-like property $\hat{c}_{a,p}\hat{c}_{a,q}\to 0$. Together, these two properties imply a certain constraint on the statistics of the $C$-variables that is ensured by special random $\theta$-numbers that are introduced in section \ref{eq: about theta numbers}.\\
\indent In addition to equations of motion (\ref{eq: stochastic equations for the particles}), appropriate initial conditions for the stochastic variables need to be defined. The initial conditions are generally not unique and only conditioned in such way, that their initial distribution recovers all possible expectation values of the system. This, in turn, is not trivial to construct for the general case. As an example in Appendix \ref{app: uncorrelated atoms}, we derive the initial conditions for a system of initially non-interacting atoms. In \cite{OLSEN20093924}, one can find examples of constructing positive P functions for various basic initial conditions. 
\subsection{$\theta$-numbers}\label{eq: about theta numbers}

In this section, we discuss the statistical properties of the $C$-variables that underlie additional constraints (Section \ref{sec: c-operators}) stemming from the fact that the general boson $c$-operators operate in a larger space of states than the original Hilbert space with the atoms. By assumption, particles cannot be annihilated, but may undergo transitions from one state to another via $\sigma$-operators. In the general language of the $c$-operators, we can, however, annihilate particles and thereby create vacuum states, or even extinguish a state by action of two consecutive annihilation operators  (\ref{eq: freedom for the bosons}). Thus, a fermion-like constraint needs to be introduced for the $C$-variables that should hold regardless of the initial conditions and the temporal evolution. Specifically, we discuss this constraint in term of statistical moments of the $C$-variables. 
Obviously, the expectation value of a single creation or annihilation operator is zero
\begin{equation*}
    \text{Tr}\left(\hat{c}_{a,p}\hat{\rho}(t)\right)=\text{Tr}\left(\hat{c}_{a,p}^\dag\hat{\rho}(t)\right)=0\;,
\end{equation*}
because action of a single creation or annihilation operator creates a vacuum state. Whereas a product of creation and annihilation operators pertaining to different atoms leads to vanishing expectation values, a product of creation and annihilation operators pertaining to the same atom gives birth to coherences $p_{a,pq}(t)$ and populations $p_{a,pp}(t)$:
\begin{multline}
\label{eq: one-particle properties}
    \langle C_{b,p}(t)C_{a,q}^\dag(t)\rangle=\text{Tr}\!\left(\hat{c}_{a,q}^\dag \hat{c}_{b,p}\hat{\rho}\!\left(t\right)\right)\\=\delta_{ab}\text{Tr}\!\left(\hat{\sigma}_{a,qp}\rho\!\left(t\right)\right)= \delta_{ab} p_{a,pq}(t).
\end{multline}
Thinking of the complex $C$-variables by a modulus and phase, equation\ (\ref{eq: one-particle properties}) suggests that each particle carries its own independent and uniformly distributed random phase. If the random phase were not compensated, this would lead to zero averages. Thus, associating a common random phase $\phi_a$ to each particle and its $C$-variables in the form of
\begin{equation*}
    C_{a,p}\sim e^{i\phi_a}, \quad C_{a,p}^\dag\sim  e^{-i\phi_a}
\end{equation*}
guarantees that all non-physical moments (not resulting from $\sigma$-operators) vanish.
Let us now find a representation of the $C$-variables that implies the fermion-like property $\hat{c}_{a,p}\hat{c}_{a,q}\to 0$. In the stochastic formalism, this property can be mimicked by special $\theta$-numbers
with the following correlators
\begin{subequations}
\begin{equation}
\label{equation: introduction of theta-numbers}
    \langle\theta^{2n}_a\rangle= \begin{cases}
    1,& \text{if }  n = 0 \text{ or } 1\\
    0,              & \text{otherwise}
\end{cases}
\end{equation}
Representing $C$-variables by the combination
\begin{equation}
\label{eq: theta-numbers}
    C_{a,p}\sim\theta_a e^{i\phi_a}, \quad C_{a,p}^\dag\sim\theta_a e^{-i\phi_a},
\end{equation}
\end{subequations}
guarantees that both imposed properties of the $C$-variables are fulfilled \footnote{The moments with odd powers correspond to an unequal number of creation and annihilation operators. Consequently, one can choose any value for such moments because the phases $\phi_a$ cancel them anyway.}.

The introduced $\theta$-numbers do not influence the expressions for the expectation values of $\hat{\sigma}_{a,pq}$ or populations and coherences $p_{a,pq}(t)$ (\ref{eq: one-particle properties}). The effect of $\theta$-numbers is more pronounced in the case of high-order moments. For instance, the expectation values for second moments lead to the following expression
\begin{multline}
\label{eq: second order correlators through C-variables}
    \text{Tr}\!\left( \hat{\sigma}_{a,pq}\hat{\sigma}_{b,rs}\hat{\rho}\!\left(t\right)\right)=\delta_{ab}\delta_{rq}\text{Tr}\!\left( \hat{c}_{a,p}^\dag\hat{c}_{a,s}\hat{\rho}\!\left(t\right)\right)\\\!\!\!\!\!\!\!\!\!+\!\text{Tr}\!\left( \hat{c}_{a,p}^\dag\hat{c}_{b,r}^\dag\hat{c}_{a,q}\hat{c}_{b,s}\hat{\rho}\!\left(t\right)\right)\!=\!\delta_{ab}\delta_{rq}\langle C_{a,p}^\dag\!\left(t\right)\!C_{a,s}\!\left(t\right)\rangle\\+\langle C_{a,p}^\dag\!\left(t\right)\!C_{b,r}^\dag\!\left(t\right)\!C_{a,q}\!\left(t\right)\!C_{b,s}\!\left(t\right)\rangle,\!\!\!\!
\end{multline}
with an additional term stemming from the normal operator ordering and the involved commutator for $c$-operators for the same atom. When $a= b$, the expression must transform into (\ref{eq: one-particle properties}). Evidently, this property is guaranteed by the representation (\ref{eq: theta-numbers}):  by construction of the $\theta$-numbers, the term containing four $C$-variables disappears and the coherences and populations are recovered by the first term. 

$\theta$-numbers are not traditional random numbers. The easiest way to sample the $\theta$-numbers is to introduce yet another phase $\theta_a=e^{i\psi_a}$, $\psi_a \in [0,2\pi)$ with a "distribution function" 
\begin{equation*}
P(\psi_a)=\frac{1+e^{-2i\psi_a}}{2\pi}.
\end{equation*}
Strictly speaking, $P(\psi_a)$ does not fulfill the requirements of a distribution function and it makes more sense to interpret $P(\psi_a)$ as a weight coefficient choosing a uniform distribution for $\psi_a$. Note, that both $C_{a,p}$ and $C_{a,p}^\dag$ get the same phase $\psi_a$.

Since $\theta$-numbers are related to one-boson Fock states, $z_a = \theta_a e^{i\phi_a}$ and $ z_a^\dag =\theta_a e^{-i\phi_a}$ can also be sampled with the following proper positive P function \cite{OLSEN20093924}
\begin{equation*}
    P(z_a^\dag,z_a)=\frac{\Big|z_a+\left(z_a^\dag\right)^*\Big|^2}{16\pi^2}\exp{\left[\left(|z_a^\dag|^2+|z_a|^2\right)/2\right]}.
\end{equation*}

Physically, the introduced random numbers do not allow two or more particles to be located at the same place, which makes them somehow similar to Grassmann numbers. Along these lines, we could interpret the $\theta$-numbers as objects pertaining to the specific algebra with $\theta^4_a=0$. Calculating expectation values $\langle...\rangle$  results in sorting out the terms with $\theta$-numbers according to correlation properties (\ref{equation: introduction of theta-numbers}).

Due to the properties of $\theta$-numbers, some of the moments vanish automatically. Consequently, we have certain freedom in defining $C$-variables (\ref{eq: theta-numbers}). In Appendix \ref{app: uncorrelated atoms}, we give an example of a system of initially non-interacting atoms. We show how the mentioned freedom allows choosing simple expressions for the initial conditions.

\section{Effective density matrix}
\label{sec: effective density}

In the previous section, we introduced $C$-variables for the particles, the stochastic analogue of creation and annihilation operators, and discussed the properties of the resulting stochastic trajectories. The $\sigma$-operators always generate pairs of creation and annihilation $c$-operators (\ref{eq: sigma operator}). In stochastic language, the resulting pairs of $C$-variables can be combined into matrix elements of an effective density matrix
\begin{equation}
    \label{eq: effective density matrix}
    \rho_{a,pq}(t)=C_{a,p}(t)C_{a,q}^\dag(t).
\end{equation}
Averaging over the stochastic trajectories of this object, indeed reproduces the elements of the one-particle density matrix, such as populations and coherences (\ref{eq: one-particle properties}).

We start the section with expressing the Hamiltonian through $\rho_{a,pq}(t)$, circumpassing the introduction of $c$-operators and their stochastic analogs. In fact, expressing the Hamiltonian through the $\sigma$-operators and replacing $\hat{\sigma}_{a,pq}$ with $\rho_{a,qp}(t)$ instantly results in the appropriate stochastic Hamiltonian function. In section \ref{subsec: stochastic equations for rho} we derive a closed system of equations for $\rho_{a,pq}(t)$. Similarly to the equations for $C$-variables, they contain derivatives of the Hamiltonian function.

The fermion-like properties of the stochastic $C$-variables that had been enforced by the $\theta$-numbers, need to be mapped onto the $\rho$-variables. To that end in section \ref{subsec: eta-numbers}, we define $\eta$-numbers for the effective density matrix defined by equation\ (\ref{equation: initial condition for the effective density matrix elements}).

Having established a concept of a density-matrix equivalent, a description of open systems becomes possible and various types of decoherence or incoherent preparation of initial states in the style of rate equations can be introduced (section \ref{sec: incoherent processes}).

In section \ref{sec: observables}, we discuss the construction of the observables in terms of the effective density matrix $\rho_{a,pq}(t)$ and give an example for the construction of initial conditions for the simple case of a system of non-interacting atoms.

\subsection{Hamiltonian}\label{sec: Hamiltonian (density)}
Before we derive the equations of motion for $\rho_{a,pq}(t)$, let us discuss the structure of the Hamiltonian and find its stochastic analogue in terms of $\rho_{a,pq}(t)$. Initially, we introduced $c$-operators to obtain simple commutation relations. Representation of the Hamiltonian by boson $c$-operators allows an unambiguous definition of normal ordering. On the other hand, their introduction extends the space of states, so that the representation of particle operators in terms of $c$-operators is not unique: normal ordered terms containing pairs of creation or annihilation operators corresponding to the same atom can be dropped (\ref{eq: freedom for the bosons}). Among all the possible ways of expanding a general operator in $c$-operators, the strategy behind equations (\ref{eq: general decomposition of an operator (c)}) seems to be the most efficient. An operator representation in the form (\ref{eq: general decomposition of an operator (sigma)}) (by analogy to boson operators, we call it normal order for $\sigma$-operators) directly generates a normal ordered decomposition in terms of boson operators (\ref{eq: general decomposition of an operator (c)}). Replacing $c$-operators with $C$-variables in this expression, we then combine them to the elements of the effective density matrix (\ref{eq: effective density matrix}). Alternatively, with the help of normal order for $\sigma$-operators (\ref{eq: general decomposition of an operator (sigma)}), we can directly substitute the $\sigma$-operators with the elements of the effective density matrix:
\begin{subequations}
\label{eq: Hamiltonian through density matrix}
\begin{equation}
    \label{eq:transformation from sigma to rho}
    \hat{\sigma}_{a,pq} \quad \to \quad \rho_{a,qp}(t)
\end{equation}
The introduction of the the $C$-variables can thus be bypassed. A general Hamiltonian $\hat{H}$ including particles and fields can thus be directly mapped to its Hamiltonian function of stochastic variables:
\begin{equation}
\label{eq: the stochastic Hamiltonian for effective density}
\begin{gathered}
    \hat{H}=\,:\!\!H\!\left(...,\hat{a}^\dag_i,\hat{a}_i,...,\hat{\sigma}_{a,qp}\right)\!\!: \\ \downarrow \\
    H(t)=H\!\left(...,\alpha^\dag_i(t),\alpha_i(t),...,\rho_{a,pq}(t)...\right).\,
\end{gathered}  
\end{equation}
\end{subequations}
Here the colons $:\!...\!:$ denote normal ordering for both boson-field operators and $\sigma$-projectors (\ref{eq: general decomposition of an operator (sigma)}). The Hamiltonian function (\ref{eq: the stochastic Hamiltonian for effective density}) is used to generate the stochastic equations of motion. Following section \ref{subsec: Hamiltonian}, we list the operator terms appearing in the Hamiltonian of the interacting system particles$+$field that result in an exact description by stochastic equations of motion:
\begin{equation*}
    \begin{split}
            \hat{\sigma}_{a,pq}\hat{\sigma}_{b,rs}, \quad\quad\hat{\sigma}_{a,pq}\hat{a}_i, \quad&\quad \hat{\sigma}_{a,pq}\hat{a}_i^\dag, \quad\quad \hat{\sigma}_{a,pq}\hat{a}_i^\dag\hat{a}_j,\\
    \big\downarrow&\\
    \rho_{a,qp}\rho_{b,sr}, \,\, \,\, \,\, \,\,\,\rho_{a,qp}\alpha_i, \,\,  \,\,\,\,\, & \,\,\,\,\, \rho_{a,qp}\alpha_i^\dag, \quad   \,\,\,\, \,\,\rho_{a,qp}\alpha_i^\dag\alpha_j.
    \end{split}
\end{equation*}

\subsection{Stochastic equation of motion\\ for the effective density matrix}
\label{subsec: stochastic equations for rho}

To get the stochastic equations for the effective density matrix (\ref{eq: effective density matrix}), we build its derivative with respect to time and recall the equation of motion for the $C$-variables (\ref{eq: equations for C-variables}). 
The stochastic equation of motion for the effective density matrix becomes
\begin{eqnarray}\label{eq: stochastic equation of motion for the effective density matrix}
        \frac{d\rho_{a,pq}}{dt}&=&\frac{i}{\hbar}\sum_r\left[\rho_{a,pr}(t)\frac{\partial H}{\partial \rho_{a,qr}}-\frac{\partial H}{\partial \rho_{a,rp}}\rho_{a,rq}(t)\right]\nonumber\\&+&\sum_r\left[ \rho_{a,pr}(t)\xi_{a,rq}^\dag\!\left(t\right)+\xi_{a,pr}\!\left(t\right)\rho_{a,rq}(t)\right],
\end{eqnarray}
where we introduce new noise terms $\xi_{a,pq}(t)$ and $ \xi_{a,pq}^\dag(t)$ with correlation functions (\ref{eq: correlation properties for the density matrix}) . Now, let us have a detailed look at the deterministic part: in the equations of motion for $C$-variables  (\ref{eq: equations for C-variables}), it was interpreted as the classical contribution to the system's evolution, as suggested by its canonical form. Similarly, the deterministic part of (\ref{eq: stochastic equation of motion for the effective density matrix}) has a well-known analogue -- the Liouville - von Neumann equation for the density matrix: First calculating the commutator $\left[\hat{\sigma}_{a,qp},\hat{H}\right]$ and then performing the transition from $\sigma$-operators to the effective density matrix $\rho_{a,pq}$ yields the deterministic part of equation\ (\ref{eq: stochastic equation of motion for the effective density matrix}).

Let us turn now to the stochastic contributions of equation\ (\ref{eq: stochastic equation of motion for the effective density matrix}). For a detailed discussion and derivation of the noise correlators, refer to appendix \ref{app: noise terms}. The stochastic variables $\xi_{a,pq}$ and $\xi\dag_{a,pq}$ have the following correlation properties:
\begin{subequations}
\label{eq: correlation properties for the density matrix}
\begin{multline}
        \langle\xi_{a,pq}^\dag\!\left(t\right)\xi_{a',p'q'}^\dag\!\left(t'\right)\rangle=-\langle\xi_{a,pq}\!\left(t\right)\xi_{a',p'q'}\!\left(t'\right)\rangle\\=\frac{i}{\hbar}\frac{\partial^2 H}{\partial \rho_{a,qp}\partial \rho_{a',q'p'}}\delta(t-t').
\end{multline}
The correlation properties of the noise terms of the bosonic field remain unchanged. Importantly, there exists a cross-correlation of the stochastic noise variables of the field and the particle system. The non-zero correltors are:
\begin{equation}
    \begin{split}
    \langle\zeta_{i}\!\left(t\right)&\xi_{a,pq}\!\left(t'\right)\rangle=-\frac{i}{\hbar}\frac{\partial^2 H}{\partial \alpha_{i}^\dag\partial \rho_{a,qp}}\delta(t-t'),\\
    \langle\zeta^\dag_{i}\!\left(t\right)&\xi_{a,pq}^\dag\!\left(t'\right)\rangle=\frac{i}{\hbar}\frac{\partial^2 H}{\partial \alpha_{i}\partial \rho_{a,qp}}\delta(t-t').
    \end{split}
\end{equation}
All other correlators vanish, in particular:
\begin{multline}
        \langle\xi_{a,pq}\!\left(t\right)\xi^\dag_{a',p'q'}\!\left(t'\right)\rangle=\langle\zeta_{i}\!\left(t\right)\xi^\dag_{a',p'q'}\!\left(t'\right)\rangle\\=\langle\zeta_{i}^\dag\!\left(t\right)\xi_{a',p'q'}\!\left(t'\right)\rangle=0.
\end{multline}
\end{subequations}
Finally, we get a closed system of equations for the effective density matrix. Equation (\ref{eq: stochastic equation of motion for the effective density matrix}) replaces equation (\ref{eq: equations for C-variables}) and the correlation properties (\ref{eq: correlation properties for the density matrix}) substitute equations (\ref{eq: correlation properties of the noise terms f}). Now remains only to specify the initial conditions. Similarly to the $C$-variables, one has to find the distribution that would recover all the possible observables for $\rho_{a,pq}(0)$. 

\subsection{$\eta$-numbers}
\label{subsec: eta-numbers}
 
As discussed in connection with the $\theta$-number -- refer to equation\ (\ref{eq: theta-numbers}) -- we need to impose fermion-like properties on $\rho_{a,pq}$. Since the effective density matrix is a product of the $C$-variables (\ref{eq: effective density matrix}), we can directly conclude on the properties of $\rho_{a,pq}$. Introducing $\eta$-numbers for squared $\theta$-numbers, we get
\begin{equation}
\label{equation: initial condition for the effective density matrix elements}
    \rho_{a,pq}\sim\eta_a ,\quad  \langle\eta^m_a\rangle= \begin{cases}
    1,& \text{if } m = 0 \text{ or } 1\\
    0,              & \text{if }m>1
\end{cases}
\end{equation}
Note there is no need of introducing a random phase $\phi_a$ due to the structure (\ref{eq: effective density matrix}) of $\rho_{a,pq}$. Like $\theta$-numbers, $\eta$-numbers are not traditional random numbers, since their dispersion is negative
\begin{equation*}
    D[\eta_a]=\langle\eta^2_a\rangle-\langle\eta_a\rangle^2=-1.
\end{equation*}
In terms of distributions, one can sample $\eta$-numbers introducing an appropriate phase $\eta_a=e^{i\psi_a}$, with $\psi_a \in [0,2\pi)$ and choosing the following "distribution function" 
\begin{equation*}
P(\psi_a)=\frac{1+e^{-i\psi_a}}{2\pi}.
\end{equation*}
The introduced random number $\psi_a$ controls the statistics of the stochastic trajectories and guarantees that terms proportional to squares of effective density matrices pertaining to the same particle vanish. In comparison to $\theta$-numbers, the similarity of $\eta$-numbers to Grassmann numbers is even more apparent, with the only difference that $\eta$-numbers commute. In simple terms, we have special algebra with $\eta^2_a=0$, and only expressions with at most one $\eta_a$ per atom survive. Calculating expectation values of observables, one just replaces non-vanishing $\eta$-numbers with one.

When the number of atoms grows, the quantum properties must average out, which in the stochastic formulation corresponds to the central limit theorem. The limit to large systems is often performed in a coarse-grained description. To that end we would define collective variables $\sum_{a}\rho_{a,pq}$, containing a sum over all atoms in a small volume. Since the atoms are close to each other and virtually indistinguishable, we assume $\rho_{a,pq}=\eta_a \times\text{const}$. With growing particle density, with the help of central limit theorem, we get
\begin{multline*}
    \sum_a \rho_{a,pq}= \text{const}\times\sum_a\eta_a\,\,\,\,\,\,\\\overset{N\to\infty}{=}\!\!\text{const}\times\left(N \mathbb{E}[\eta]+\sqrt{ND[\eta]}\kappa+...\right)\\=\text{const}\times\left(N+i\sqrt{N}\kappa+...\right)
\end{multline*}
where $\mathbb{E}[\eta]=1$ and $D[\eta]=-1$ are the expectation value and variance of $\eta_a$. The random number $\kappa$ has a normal distribution with zero mean and variance equal one. Assuming a very large number of atoms $N$, one can neglect the random part. Consequently, only the deterministic part remains, which means that the non-trivial random $\eta$-numbers disappear. The same holds for $C$-variables: if the density of the atoms is high, one can neglect $\theta$-numbers. 

In addition to the proposed approaches to $\eta$-numbers, one can average them before solving the differential equations. This can be implemented in the so called gauging techniques that are discussed in section  \ref{sec: stochastic gauges}.
\subsection{Observables}
\label{sec: observables}
Having introduced appropriate stochastic variables and their evolution, we now express expectation values in terms of the effective density matrix $\rho_{a,pq}(t)$. We resort to the $\sigma$-representation of single-particle operators (\ref{eq: operator decomposition}). Thus, every single-particle expectation value links to the expectation value of the corresponding $\sigma$-projector. In a bulky, but straightforward derivation we find that the quantum-mechanical average of the $\sigma$-projector equals the ensemble average over stochastic trajectories of the effective density matrix
\begin{subequations}
\label{expectation values x particle}
\begin{equation}
    \label{expectation values one particle}
    \langle\rho_{a,pq}\!\left(t\right)\rangle=\text{Tr}\!\left( \hat{\sigma}_{a,qp}\hat{\rho}\!\left(t\right)\right).
\end{equation}
Two-particle observables are linked to second-order moments of the $\sigma$-operator (pertaining to different particles), which can be sampled by:
\begin{multline}
    \label{expectation values two particle}
    \langle \rho_{a,pq}\!\left(t\right)\rho_{b,rs}\!\left(t\right)\rangle\\=\begin{cases}
    \text{Tr}\!\left( \hat{\sigma}_{a,pq}\hat{\sigma}_{b,rs}\hat{\rho}\!\left(t\right)\right),& \text{if }  a \neq b\\
    0,              & \text{otherwise}
\end{cases}
\end{multline}
For $a=b$, the result is zero, ensured by the $\eta$-numbers. For higher-order moments, describing $n$-particle interactions of distinct particles, equation (\ref{expectation values two particle}) generalizes to 
\begin{multline}
    \langle {\rho}_{a,qp}\!\left(t\right)\rho_{b,sr}\!\left(t\right)...\rho_{d,ji}\!\left(t\right)\rangle\\= \text{Tr}\!\left( \hat{\sigma}_{a,pq}\hat{\sigma}_{b,rs}...\hat{\sigma}_{d,ij}\hat{\rho}\!\left(t\right)\right),
\end{multline}
\end{subequations}
with the understanding that for any pair of indices of $a,b,...d,...$ being identical, the expectation value is zero. 
In other respects, the resulting expressions follow the inverse map (\ref{eq: Hamiltonian through density matrix}) 
\begin{equation*}
    \hat{\sigma}_{a,pq} \quad \xleftarrow{} \quad \rho_{a,qp}(t).
\end{equation*}  
Using equations (\ref{expectation values x particle}), one can construct the initial distributions for the elements of the effective density matrices. For example, if one considers a system of initially non-interacting atoms, the traces in the expressions (\ref{expectation values x particle}) decouple into the products of the initial one-particle density matrices $p_{a,pq}^{(0)}$:
\begin{equation*}
    \text{Tr}\!\left( \hat{\sigma}_{a,pq}\hat{\sigma}_{b,rs}...\hat{\sigma}_{d,ij}\hat{\rho}\!\left(0\right)\right)=p_{a,pq}^{(0)}p_{b,rs}^{(0)}...p_{d,ij}^{(0)}
\end{equation*}
Consequently, a simple choice for appropriate initial conditions is $\rho_{a,pq}(t=0)=\eta_a p_{a,pq}^{(0)}$.

\subsection{Incoherent processes}
\label{sec: incoherent processes}

Closed physical systems are rare. Typically, a physical systems interacts with an "outer world", sometimes causing significant losses and decoherence in the considered physical subsystem. In the operator formalism, the analysis of such open systems is typically performed with quantum-Langevin equations \cite{drummond2014quantum, scully1999quantum} or with Lindblad master equations \cite{Breuer2007, lindblad1976generators, gorini1976completely, agarwal1974quantum}
\begin{multline}
    \label{eq: master equation with dicoherence}
    \frac{d\hat{\rho}\!\left(t\right)}{dt}=-\frac{i}{\hbar}\left[\hat{H},\hat{\rho}\!\left(t\right)\right]\\+\frac{1}{2}\sum_{i}\left(\left[\hat{L}_i\hat{\rho}(t),\hat{L}_i^\dag\right] +\left[\hat{L}_i,\hat{\rho}(t)\hat{L}_i^\dag\right]\right),
\end{multline}
where the additional terms preserve the trace of the density matrix and keep it always hermitian and positive semi-definite. Operators $\hat{L}_i$ are called the Lindblad or jump operators.

Often, the evolution of atomic systems is strongly governed by incoherent processes such as non-radiative decay, ionization or decoherence \cite{Meystre2007, PhysRevA.99.013839, briegel1993quantum}. Typically, these processes are assumed to be independent for each atom. Consequently, we can associate each atom with a separate reservoir. In the following, we omit index $a$ that enumerates the atoms. The most general master equation for the considered scenario has the following structure
\begin{multline}
\label{eq: master equation with incoherences}
    \frac{d\hat{\rho}\!\left(t\right)}{dt}=\frac{i}{\hbar}\left[\hat{\rho}\!\left(t\right),\hat{H}\right]\\+\!\frac{1}{2}\!\!\sum_{p,q,r,s}\!\!\!\Gamma_{pqrs}\!\left(t\right)\left(\left[\hat{\sigma}_{pq}\hat{\rho}\!\left(t\right),\hat{\sigma}_{sr}\right]+\left[\hat{\sigma}_{pq},\hat{\rho}\!\left(t\right)\hat{\sigma}_{sr}\right]\right),\!\!\!\!
\end{multline}
where $\Gamma_{pqrs}^*\!\left(t\right)=\Gamma_{rspq}\!\left(t\right)$. In terms of one-particle density matrix elements $p_{ij}\!\left(t\right)=\textrm{Tr}\!\left(\hat{\sigma}_{ji}\hat\rho\!\left(t\right)\right)$, one gets so-called rate equations
\begin{multline}
\label{eq: rate equations}
    \!\!\left(\frac{dp_{pq}\!\left(t\right)}{dt}\right)_{\textrm{non-unitary}}\!\!\!\!\!\!=\!\frac{1}{2}\sum_{r,s}\Big(2\Gamma_{prqs}p_{rs}(t)\\-\Gamma_{rqrs}p_{ps}(t)-\Gamma_{rsrp}p_{sq}(t)\Big).\!\!
\end{multline}

In Appendix~\ref{app: incoherences}, we introduce the corresponding Hamiltonian (\ref{eq: updated Hamiltonian}) and the detailed derivations of the corresponding stochastic equations. The derivations lead to the following additional terms in the stochastic equations for the effective density matrix elements 
\begin{multline}
\label{eq: rate equations 2}
    \frac{d\rho_{a,pq}\!\left(t\right)}{dt}=...+\!\frac{1}{2}\sum_{r,s}\Big(2\Gamma_{a,prqs}\rho_{a,rs}(t)\\-\Gamma_{a,rqrs}\rho_{a,ps}(t)-\Gamma_{a,rsrp}\rho_{a,sq}(t)\Big).
\end{multline}
 Note, that the resulting equation has no additional noise terms. The new terms, describing coupling to the bath, are identical to rate equations (\ref{eq: rate equations}). The interpretation of $\rho_{a,ij}$ as an effective density matrix is thus also fully justified for open quantum systems.

\section{STOCHASTIC GAUGES}
\label{sec: stochastic gauges}

The transition from the operator formalism to the stochastic sampling reveals a lot of freedom in the way, the phase space is sampled. Thus, stochastic trajectories and the corresponding system of equations of motions are not uniquely defined. This relates to the fact, that different probability distributions can generate the same moments \cite{Romano2017}. We note that the most obvious choice of the stochastic equation does not necessarily lead to a stable numerical solution.  The stability issues motivated the development of so-called stochastic gauges \cite{2002'Deuar_gauge,2003'Drummond_gauge,2006'Deuar_gauge,2005'DeuarPhD}. In \cite{2005'DeuarPhD}, the derivation of the stochastic gauge transformation was based on the freedom in decomposing the density matrix in terms of $\Lambda$-operators (\ref{eq: density matrix decomposition}): on the one hand, one can choose different $\Lambda$-operators; on the other hand, $\Lambda$-operators are analytical functions of their arguments, which provides even more freedom. All these observations indicate that the $P$-distribution is not unique. Since it is sampled with the stochastic trajectories, their choice is not unique as well.

In our formalism we effectively circumvent the $P$-distribution, the stochastic trajectories being the key objects of our approach. Besides, in section \ref{sec: effective density}, we introduce the effective density matrix, which has no direct link to the $\Lambda$-projectors of the known approaches of stochastic gauge transformation. This motivates us to introduce the stochastic-gauge freedom in a more general context. \\
\indent Let us start by investigating a general stochastic process $\textbf{x}(t)$. The exact form of the stochastic equations of motion and their origin may be disregarded. Consider a so-called characteristic function
\begin{equation*}
    \chi(\bm{\lambda},t)=\Big\langle \exp{\big[\bm{\lambda}\cdot\textbf{x}(t)\big]}\Big\rangle.
\end{equation*}
Its derivatives give all the necessary information to calculate moments and any expectation values of any observable $f[\textbf{x}(t)]$: 
\begin{equation*}
    \langle f[\textbf{x}(t)]\rangle = f\!\left[\frac{\partial}{\partial  \bm{\lambda}}\right]\chi(\bm{\lambda},t)\Big|_{\bm{\lambda}=\textbf{0}}.
\end{equation*}
Consequently, $\chi(\bm{\lambda},t)$ is uniquely defined in the vicinity of $\bm{\lambda}=\textbf{0}$, since its derivatives at the point $\bm{\lambda}=\textbf{0}$ determine all observables.\\
\indent Now, let us try to construct a different system of equations that lead to a new set of stochastic trajectories $\textbf{x}'(t)$. One can find a corresponding characteristic function $\chi'(\bm{\lambda},t)$. To be related to the same physical system, $\chi(\bm{\lambda},t)$ and $\chi'(\bm{\lambda},t)$ must generate the same moments and expectation values of all possible observables. Because of their defining property, the new and old characteristic functions must be equivalent:
\begin{equation}
\label{eq: condition for chi}
    \chi(\bm{\lambda},t)=\chi'(\bm{\lambda},t).
\end{equation}
Usually, an explicit form of the stochastic trajectories $\textbf{x}(t)$ or $\textbf{x}'(t)$ is unknown, only their equations of motion are given, so we cannot at once build the corresponding characteristic functions and compare them. We can only act in the spirit of the mathematical induction technique, to map an ensemble  of stochastic trajectory $\textbf{x}(t)$ to a statistically equivalent ensemble $\textbf{x}'(t)$.  First, one assures that the initial conditions for $\textbf{x}'(0)$ recover the same characteristic function, i.e.\ satisfying (\ref{eq: condition for chi}) for $t=0$. Then, assuming that (\ref{eq: condition for chi}) holds for later times $t$, one ensures that the temporal derivatives of the charactersitic function are preserved:
\begin{eqnarray}\label{eq: condition for chi 2}
    \frac{\partial}{\partial t} \chi(\bm{\lambda},t)&=&\frac{\partial}{\partial t}\chi'(\bm{\lambda},t)
\end{eqnarray}

This is the key equation, defining the {\it stochastic gauge freedom} for the stochastic trajectories. In appendix \ref{app: drift gauge}, we reproduce Girsanov theorem \cite{liptser2001statistics} and a drift gauge discussed thoroughly in  \cite{2005'DeuarPhD}. 

Besides, equation\ (\ref{eq: condition for chi 2}) allows complete removal of $\eta$-numbers that possess inconvenient statistical properties. Here, we limit ourselves only with the basic idea and leave the details for an upcoming publication. Assume that one single $\eta$-number enters the initial conditions $\textbf{x}(0)=\eta\times\textbf{x}_0$.  The key step is to build a new characteristic function that takes explicitly into account the presence of $\eta$-numbers
\begin{equation*}
    \chi'(\bm{\lambda},t)=\Big\langle \exp{\big[\eta\bm{\lambda}\cdot\textbf{x}'(t)\big]}\Big\rangle=\Big\langle 1+\bm{\lambda}\cdot\textbf{x}'(t)\Big\rangle,
\end{equation*}
where $\textbf{x}'(t)$ are new trajectories with the initial conditions without $\eta$-numbers $\textbf{x}(0)=\textbf{x}_0$. It turns out that for many problems one can build such new trajectories, so the resulting equations operate with the variables possessing intuitive complex-valued initial conditions.

\section{Discussion and Outlook}
\label{sec: discussion}

After presenting the key ingredients of our formalism, we suggest a list of specific problems that can be straightforwardly translated into the stochastic language. 

\subsection{Light-matter interaction}

A wide variety of phenomena involving light-matter interaction satisfy the conditions for the stochastic interpretation. The typical minimal-coupling interaction Hamiltonian 
\begin{equation*}
    \label{eq: minimal coupling hamiltonian}
    \sum_{a;i,p,q}\hbar \left(g_{i;a,p,q}\hat{a}_i+g_{i;a,q,p}^*\hat{a}_i^\dag\right)\hat{\sigma}_{a,pq}
\end{equation*}
only accounts for pair interaction between atoms and fields. The evolution of atomic systems is usually influenced by incoherent processes \cite{Meystre2007, PhysRevA.99.013839, briegel1993quantum} that can be treated independently for each atom. At the level of the effective density matrix, it allows for equation (\ref{eq: rate equations 2}) that does not include any additional noise terms. 

In many problems of light-matter interaction, the classical behaviour dominates the quantum one. In particular, such a many-body effect as collective spontaneous emission is typically seen as a classical process enriched by a phenomenological noise term or randomness in the initial conditions \cite{2000'Larroche,2014'Weninger,2018'Krusic,2020'Lyu, gross1982superradiance, 1979'Haake,1979'Vrehen, PhysRevA.101.023836}. In our formalism, the noise term appears naturally without the need of any ad hoc approximations. In the context of light-matter interaction, the deterministic part of equation\ (\ref{eq: stochastic equation of motion for the effective density matrix}) corresponds to the optical Bloch equations, whereas the noise contribution enriches these equations with the spontaneous nature of the quantum mechanics allowing a correct treatment of the spontaneous emission. 

In a forthcoming publication, our focus is on light-matter interaction in free space. In particular, we will present a rigorous description of superradiace and collective spontaneous emission in realistic conditions and a broade range of physical parameters. 

Another important family of models are various atomic systems in cavities \cite{10.1088/978-0-7503-3447-1}. In the Tavis-Cummings model, the well-known phenomenon of collapse and revival \cite{ scully1999quantum} can be seen as the interference of different realisations of classical Rabi oscillations distorted by quantum noise terms. In contrast to the free-space models, a reduced number of modes leads to a stronger memory effect, which may turn out inherently difficult to reproduce with delta-correlated stochastic processes. For example in \cite{Mandt_2015}, the positive P representation for spin+boson interaction based on coherent states has been benchmarked against the Monte Carlo wave-function approach \cite{Molmer:93}, which shows good agreement only in the steady state regime.

\subsection{Electronic-structure theory and quantum-chemistry approaches}
Many attempts have been made to describe electronic dynamics inside atoms by means of phase-space methods \cite{DALTON201612, DALTON2017268, Polyakov2016, PhysRevA.64.063409, Kidwani_2020}. A typical Hamiltonian for electronic degrees of freedom \cite{szabo1982modern} reads
\begin{equation}
\label{eq: quantum chemistry hamiltonian}
\hat{H}_{\text{QCh}}=\sum_{p,q}F_{pq}\hat{a}_p^\dag\hat{a}_q +\frac{1}{2}\sum_{p,q,r,s}W_{pqrs}\hat{a}_p^\dag\hat{a}_q^\dag\hat{a}_s\hat{a}_r.
\end{equation}
Here, the first term encompasses the kinetic energy and nuclear-electron attraction, whereas the second term represents the electron-electron repulsion in second quantization. Since our approach is independent of the statistics of particles, one can straightforwardly apply the derived formalism to eq. (\ref{eq: quantum chemistry hamiltonian}) expressed through $\sigma$-operators
\begin{equation*}
\hat{H}_{\text{QCh}}=\sum_{a,p,q}F_{pq}\hat{\sigma}_{a,pq} +\sum_{p,q,r,s}W_{pqrs}\sum_{a> b}\hat{\sigma}_{a,pr}\hat{\sigma}_{b,qs}.
\end{equation*}
Since the second term contains products of $\sigma$-operators, electron-electron repulsion is responsible for the presence of the noise terms in the final stochastic equations. This contrasts to the basic idea of Quantum Monte Carlo  \cite{Austin2012-wn} where the diffusion comes from the kinetic energy. 

Although formally different, the resulting stochastic equations will be qualitatively similar to other fermionic phase-space approaches \cite{DALTON201612, DALTON2017268, Polyakov2016, PhysRevA.64.063409}. Our formalism, however, allows the unified treatment of particles with different statistic, which allows us to include also nuclear degrees of freedom and derive comprehensive quantum chemistry methods beyond the Born-Oppenheimer approximation. This can, in turn, open prospects to a theoretical and numerical analysis of dynamics around conical intersections based on decoupled stochastic differential equations for electronic and nuclear degrees of freedom. The interaction part of the Hamiltonian includes the coupling between nuclei and electrons via
\begin{equation*}
\sum_{p,q,r}U_{pqr}\sum_{a,B}\hat{\sigma}_{a,pq}\hat{\sigma}_{B,rr}, 
\end{equation*}
which gives birth to additional noise terms. Here, capital letter $B$ stands for nuclei and $r$ represents state $\ket{\textbf{r}}$ in the position basis. The new noise terms bring uncertainty to the nuclear dynamics and can be seen as an alternative to other existing statistical methods \cite{doi:10.1063/1.459170, doi:10.1021/ct1007394}.

\begin{acknowledgments}
We would like to thank Dr. Aliaksandr Leonau for the constructive and critical discussion of the manuscript. We acknowledge the financial support of Grant-No.~HIDSS-0002 DASHH (Data Science in Hamburg-Helmholtz Graduate School for the Structure of Matter).  We are grateful to Mathias Trabs for extended discussions on stochastic differential equations and pointing us towards the Girsanov theorem.
\end{acknowledgments}

\bibliography{bib}

\appendix

\section{Derivation of $\mathcal{L}$}
\label{app: derivation of L}
We start with the master equation (\ref{eq: initial master equation}) and use representation (\ref{eq: density matrix through the trajectories}) for the density matrix
\begin{equation*}
    \frac{\partial}{\partial t}\Big\langle\hat{\Lambda}\!\left(\alpha(t),\alpha^\dag(t)\right)\Big\rangle=-\frac{i}{\hbar}\Big\langle\left[\hat{H},\hat{\Lambda}\!\left(\alpha(t),\alpha^\dag(t)\right)\right]\Big\rangle.
\end{equation*}

Thanks to the normal order, the commutator in the resulting expression has a simple form. First, we can apply the first property from equation (\ref{eq: properties of lambda})
\begin{multline*}
    \left[\,:\!\!H\!\left(\hat{a}^\dag,\hat{a}\right)\!\!:,\hat{\Lambda}(\alpha,\alpha^\dag) \right]\\=H\!\left(\hat{a}^\dag,\alpha\right)\hat{\Lambda}(\alpha,\alpha^\dag)-\hat{\Lambda}(\alpha,\alpha^\dag)H\!\left(\alpha^\dag,\hat{a}\right)
\end{multline*}
Since the Hamiltonian contains now only complex numbers and commuting operators we do not need the normal order anymore. The next step is to use the second property from equation (\ref{eq: properties of lambda}). Note, that this property replaces operators with derivatives acting \textbf{only} on the $\Lambda$-operators and we have to be careful with the order of the stochastic variables and the respective derivatives inside the Hamiltonian
\begin{multline*}
    =\,:\!\!\Bigg[H\!\left(\alpha^\dag+\frac{\partial}{\partial \alpha},\alpha\right)-H\!\left(\alpha^\dag,\alpha+\frac{\partial}{\partial \alpha^\dag}\right)\Bigg]\!\!:\\\times\hat{\Lambda}(\alpha,\alpha^\dag),
\end{multline*}
where $:...:$ is used for the normal order: all the derivatives go to the right. Consequently, the derivatives act only on the $\Lambda$-operators. Finally, we get
\begin{equation*}
    \frac{\partial}{\partial t}\Big\langle\hat{\Lambda}\!\left(\alpha(t),\alpha^\dag(t)\right)\Big\rangle=\frac{i}{\hbar}\Big\langle \mathcal{L}\,\hat{\Lambda}\!\left(\alpha(t),\alpha^\dag(t)\right)\Big\rangle.
\end{equation*}
where we introduce an operator $\mathcal{L}$ to shorten the equation
\begin{equation*}
    \mathcal{L}=\,:\!H\!\left(\alpha^\dag,\alpha+\frac{\partial}{\partial \alpha^\dag}\right)\!:-:\!H\!\left(\alpha^\dag+\frac{\partial}{\partial \alpha},\alpha\right)\!:\,.
\end{equation*}
It is more practical to decompose $\mathcal{L}$ into a Taylor series. Equation (\ref{eq: decomposition of L}) contains its second Taylor polynomial that can be sampled by the stochastic trajectories.
\section{Ito's lemma}\label{app: ito}
Besides deterministic contributions, equations for the stochastic field variables
\begin{align*}
\frac{d\alpha\!\left(t\right)}{dt}\!=&A(t)+\zeta\!\left(t\right),\\
\frac{d\alpha^\dag\!\left(t\right)}{dt}=&A^\dag(t)+\zeta^\dag\!\left(t\right)
\end{align*}
include the Gaussian complex white noise terms $\zeta\!\left(t\right)$ and $\zeta^\dag\!\left(t\right)$ with the following correlation properties 
\begin{align*}
\langle \zeta\!\left(t\right)\zeta\!\left(t'\right)\rangle=&\,\sigma^{(-)}(t)\delta(t-t'),\\
\langle \zeta^\dag\!\left(t\right)\zeta^\dag\!\left(t'\right)\rangle=&\,\sigma^{(+)}(t)\delta(t-t'),\\
\langle \zeta\!\left(t\right)\zeta^\dag\!\left(t'\right)\rangle=&\,\sigma^{(0)}(t)\delta(t-t').
\end{align*}
Due to stochastic nature, the chain rule for differentiation of the functions of the stochastic processes requires a revision. The corrected chain rule for the stochastic processes in the Ito definition is called Ito's lemma \cite{gardiner2004handbook}. For an analytical function $f\!\left(\alpha(t),\alpha^\dag(t)\right)$ of the stochastic processes $\alpha\!\left(t\right)$ and $\alpha^\dag\!\left(t\right)$, the differentiation leads not only to the traditional linear components, but also to terms proportional to the diffusion coefficients:
\begin{widetext}
\begin{multline*}
\frac{d}{dt}f\!\left(\alpha(t),\alpha^\dag(t)\right)=\Bigg[\left(A^\dag(t)+\zeta\!\left(t\right)\right)\frac{\partial}{\partial \alpha^\dag}+\left(A(t)+\zeta^\dag\!\left(t\right)\right)\frac{\partial}{\partial \alpha}\\+\frac{\sigma^{(-)}(t)}{2}\frac{\partial^2}{\partial \alpha^2}+\frac{\sigma^{(+)}(t)}{2}\frac{\partial^2}{\partial \alpha^{\dag 2}}+\sigma^{(0)}(t)\frac{\partial^2}{\partial \alpha^{\dag}\partial\alpha}\Bigg]\,f\!\left(\alpha(t),\alpha^\dag(t)\right).
\end{multline*}
\end{widetext}

Our task is to find the form of the deterministic drift coefficients $A(t)$, $A^\dag(t)$ and diffusion coefficients $\sigma^{(-)}(t)$, $\sigma^{(+)}(t)$ and $\sigma^{(0)}(t)$ that would recover the master equations (\ref{eq: equation for the averaged lambda}). Since $\Lambda$-projectors are analytical functions of their arguments, we can replace $f\!\left(\alpha(t),\alpha^\dag(t)\right)$ with $\hat{\Lambda}\!\left(\alpha(t),\alpha^\dag(t)\right)$. The next step is to average the resulting equation, which cancels the contributions containing the Ito noise terms, because they can only correlate with the variables from the future
\begin{multline*}
    \frac{d}{dt}\Big\langle
    \hat{\Lambda}\!\left(\alpha(t),\alpha^\dag(t)\right)\!\Big\rangle=\Big\langle \bigg[A^\dag(t)\frac{\partial}{\partial \alpha^\dag}+A(t)\frac{\partial}{\partial \alpha}\\+\frac{\sigma^{(-)}(t)}{2}\frac{\partial^2}{\partial \alpha^2}+\frac{\sigma^{(+)}(t)}{2}\frac{\partial^2}{\partial \alpha^{\dag 2}}\\+\sigma^{(0)}(t)\frac{\partial^2}{\partial \alpha^{\dag}\partial\alpha}\bigg]\,
    \hat{\Lambda}\!\left(\alpha(t),\alpha^\dag(t)\right)\!\Big\rangle,
\end{multline*}

The stochastic trajectories recover the second order derivatives in equation (\ref{eq: equation for the averaged lambda}), if one chooses the following diffusion coefficients  
\begin{equation*}
       \sigma^{(-)}(t)=-\frac{i}{\hbar}\frac{\partial^2 H}{\partial \alpha^{\dag 2}},\,\,\,\,\,\,
        \sigma^{(+)}(t)=\frac{i}{\hbar}\frac{\partial^2 H}{\partial \alpha^2},\,\,\,\,\,\,
        \sigma^{(0)}(t)=0,
\end{equation*}
that lead to the correlation properties of equations\ (\ref{eq: correlation properties for two bosons}).

\section{Examples of the bosonization}
\label{app: examples of the bosonization}

\subsection{Bosonization of atoms}
As simple example of a quantum particle we consider an atom represented by finite amount of electronic levels. Consider two independent non-interacting atoms in stationary states $\ket{p}$ and $\ket{q}$ respectively
\begin{equation*}
    \ket{\Psi}=\ket{p}_1\otimes\ket{q}_2,
\end{equation*}
where integers 1 and 2 distinguish the particles. Introducing creation and annihilation bosonic operators for each atom
\begin{align*}
    \hat{c}^\dag_{1,p}, \,\, \hat{c}_{1,p}, \,\, \hat{c}^\dag_{2,q}, \,\, \hat{c}_{2,q},
\end{align*}
that $\ket{\Psi}$ can be constructed by
\begin{equation*}
    \ket{\Psi}=\hat{c}^\dag_{1,p}\hat{c}^\dag_{2,q}\ket{\o},
\end{equation*}
where $\ket{\o}$ is an effective vacuum state. A general quantum state is thus represented by
\begin{equation*}
    \ket{\Psi}=\sum_{p,q}C_{pq}\hat{c}^\dag_{1,p}\hat{c}^\dag_{2,q}\ket{\o}.
\end{equation*}
Let us now have a look at the Hamiltonian. The free Hamiltonian of two independent atoms has the following form
\begin{equation*}
    \hat{H}=\sum_{p}\varepsilon_p\left(\hat{\sigma}_{1,pp}+\hat{\sigma}_{2,pp}\right)=\sum_{p}\varepsilon_p\left(\hat{c}^\dag_{1,p}\hat{c}_{1,p}+\hat{c}^\dag_{2,p}\hat{c}_{2,p}\right),
\end{equation*}
where $\varepsilon_p$ is the energy of the stationary state $\ket{p}$. One may wish to add interaction between this atoms. A simple and illustrative example is the dipole-dipole interaction
\begin{equation*}
    \hat{V}=\frac{1}{4\pi \varepsilon_0 r^3}\,\hat{\textbf{d}}_1\!\left(1-3\frac{\textbf{r}\otimes\textbf{r}}{r^2}\right)\hat{\textbf{d}}_2,
\end{equation*}
where $\textbf{r}$ with $|\textbf{r}|=r$ is a vector pointing from atom 1 to 2, $\hat{\textbf{d}}_1$ and $\hat{\textbf{d}}_2$ are the dipole moment operators of atom 1 and 2, respectively. In terms of the $c$-operators, this expression reads
\begin{equation*}
    \hat{V}=\sum_{p,q,r,s}\frac{\hat{c}^\dag_{1,p}\hat{c}_{1,q}\hat{c}^\dag_{2,r}\hat{c}_{2,s}}{4\pi \varepsilon_0 r^3}\,\textbf{d}_{pq}\!\left(1-3\frac{\textbf{r}\otimes\textbf{r}}{r^2}\right)\textbf{d}_{rs},
\end{equation*}
where $\textbf{d}_{pq}=\braket{p|\textbf{d}|q}$ are the dipole moment matrix elements.

\subsection{Bosonisation of fermions}
Note, that in our formalism we do not specify the statistical properties of the quantum particles,  they can be fermions or bosons. Consider the case of two indistinguishable non-interacting fermions enumerated with $1,2$ in states $\ket{p}$ and $\ket{q}$
\begin{equation*}
    \ket{\Psi}=\frac{1}{\sqrt{2}}\big(\ket{p}_1\otimes\ket{q}_2-\ket{p}_2\otimes\ket{q}_1\big).
\end{equation*}
Introducing bosonic creation and annihilation bosonic operators for each fermion
\begin{align*}
    \hat{c}^\dag_{1,p}, \,\, \hat{c}_{1,p}, \,\, \hat{c}^\dag_{2,p}, \,\, \hat{c}_{2,p},\\\hat{c}^\dag_{1,q}, \,\, \hat{c}_{1,q}, \,\, \hat{c}^\dag_{2,q}, \,\, \hat{c}_{2,q},
\end{align*}
allows us to generate the states from an effective vacuum state $\ket{\o}$
\begin{align*}
    \ket{p}_1 = \hat{c}^\dag_{1,p}\ket{\o}, \quad \ket{q}_1 = \hat{c}^\dag_{1,q}\ket{\o}, \\
    \ket{p}_2 = \hat{c}^\dag_{2,p}\ket{\o}, \quad \ket{q}_2 = \hat{c}^\dag_{2,q}\ket{\o}.
\end{align*}
The state $\ket{\Psi}$ is thus represented by
\begin{equation*}
    \ket{\Psi}=\frac{1}{\sqrt{2}}\big( \hat{c}^\dag_{1,p}\hat{c}^\dag_{2,q}-\hat{c}^\dag_{2,p}\hat{c}^\dag_{1,q}\big)\ket{\o}.
\end{equation*}
Note, that Pauli principle and anti-symmetrical properties are fully recovered.

In a more general case of $N$ indistinguishable non-interacting fermions, the state of the system is a Slater determinant
\begin{align*}
    \ket{\Psi}=\frac{1}{\sqrt{N!}}\det\begin{pmatrix} \ket{1}_1 & \ket{2}_1 & \vdots & \ket{N}_1\,\,\, \\ \ket{1}_2 & \ket{2}_2 & \vdots & \ket{N}_2\,\,\, \\ \dots &\dots &\ddots &\dots \,\,\,\\ \ket{1}_N & \ket{2}_N & \vdots & \ket{N}_N\,\,\end{pmatrix},
\end{align*}
In terms of $c$-operators, $\ket{\Psi}$ transforms into the following expression 
\begin{align*}
    \ket{\Psi}=\frac{1}{\sqrt{N!}}\det\begin{pmatrix} \hat{c}^\dag_{1,1} & \hat{c}^\dag_{1,2} & \vdots & \hat{c}^\dag_{1,N}\,\,\, \\ \hat{c}^\dag_{2,1} & \hat{c}^\dag_{2,2} & \vdots &  \hat{c}^\dag_{2,N} \,\,\, \\ \dots &\dots &\ddots &\dots \,\,\,\\  \hat{c}^\dag_{1,N}  & \hat{c}^\dag_{2,N} & \vdots & \hat{c}^\dag_{N,N}\,\,\end{pmatrix}\ket{\o},
\end{align*}
Other strategies for bosonization of fermions can be found for example in reference \cite{MontoyaCastillo2018}.

\section{A system of uncorrelated atoms}
\label{app: uncorrelated atoms}
An indispensable ingredient of the stochastic phase-space formalism is the definition of the initial conditions for the stochastic trajectories $\textbf{C}(0)$, $\textbf{C}^\dag(0)$, $\bm{\alpha}(0)$, $\bm{\alpha}^\dag(0)$. The initial distribution from which starting points of the trajectories are sampled, is not unique and only constraint by the expectation values of all possible observables to match those of the initial quantum state. One can thus reshape this initial distribution and make any choice suited for a particular problem.

As the title of the section suggests, let us consider a system of initially uncorrelated atoms. In this case, it is sufficient to look at the initial conditions for an arbitrary atom $a$ with a general initial wave function
\begin{equation*}
    \label{eq: pure state}
    \ket{\Psi\!\left(0\right)}_a=\sum_pC^{(0)}_{a,p}\ket{p}_a.
\end{equation*}
Here, the state of the atom is decomposed in terms of some orthogonal basis set $\{\ket{p}\}$. Let us construct various possible expectation values, involving $C_{a,p}(0)$ and $C^\dag_{a,q}(0)$. $\langle C_{a,p}(0)C^\dag_{a,q}(0)\rangle$ correspond to the populations and coherences of an atom 
\begin{multline*}
    \langle C_{a,p}(0)C^\dag_{a,q}(0)\rangle=\text{Tr}\!\left(\hat{c}_{a,q}^\dag\hat{c}_{a,p}\hat{\rho}\!\left(0\right)\right)\\=\text{Tr}\!\left(\hat{\sigma}_{a,qp}\hat{\rho}\!\left(0\right)\right)=C^{(0)}_{a,p}C^{(0)*}_{a,q}.
\end{multline*}
All the higher orders equal zero
\begin{multline*}
    \langle C_{a,p}(0)C_{a,r}(0)C^\dag_{a,q}(0)C^\dag_{a,s}(0)\rangle\\=\text{Tr}\!\left(\hat{c}_{a,q}^\dag\hat{c}_{a,s}^\dag\hat{c}_{a,p}\hat{c}_{a,r}\hat{\rho}\!\left(0\right)\right)=0,
\end{multline*}
which is automatically recovered by $\theta$-numbers. Consequently, we only have to satisfy the expressions for the first moments accompanying $\theta$-numbers (\ref{eq: theta-numbers}) with the usual decomposition coefficients $C^{(0)}_{a,p}$ (\ref{eq: pure state})
\begin{equation*}
\begin{split}
    C_{a,p}\!\left(0\right)&=\theta_aC^{(0)}_{a,p}e^{i\phi_a},\\C^\dag_{a,p}\!\left(0\right)&=\theta_aC^{(0)*}_{a,p}e^{-i\phi_a}.
\end{split}
\end{equation*}
Finally, the coefficients of the decomposition of the wave function enter naturally the initial conditions for the $C$-variables. The quantum properties are recovered by $\theta_a$ and uniformly distributed phases $\phi_a$.

\section{Noise terms in the equations for the effective density matrix}
\label{app: noise terms}

In section \ref{subsec: stochastic equations for rho}, we introduce the equation of motion for the effective density matrix. Its derivation starts with differentiating equation (\ref{eq: effective density matrix}) with respect to time. Next, we have to use the equation of motion for the $C$-variables (\ref{eq: equations for C-variables}). The resulting equation has the following structure
\begin{equation*}
        \frac{d\rho_{a,pq}\!\left(t\right)}{dt}=...+F_{a,pq}(t)+F_{a,pq}^\dag(t),
\end{equation*}
where we drop the deterministic part and collect the noise terms into $F_{a,pq}(t)$ and $F_{a,pq}^\dag(t)$
\begin{equation}
\label{eq: F noise terms}
    \begin{split}
       F_{a,pq}(t)=\xi_{a,p}\!\left(t\right)C_{a,q}^\dag(t),\\
       F_{a,pq}^\dag(t)=\xi_{a,q}^\dag\!\left(t\right)C_{a,p}(t),
    \end{split}
\end{equation}
that function as noise terms as well. At first glance, $C_{a,p}(t)$ and $C_{a,q}^\dag(t)$ do not pair into the density matrix elements. However, we have an additional freedom: only the correlation properties between the involved noise terms matter. Consequently, we can always choose any explicit form for the noise terms that reproduce the true correlators. Let us determine first the correlators between  $F_{a,pq}(t)$, $F_{a,pq}^\dag(t)$ and the field noise terms. Using definition (\ref{eq: F noise terms}) and properties of the old noise terms (\ref{eq: correlation properties of the noise terms f}) and (\ref{eq: correlation properties of the noise terms f (ad)}), one gets
\begin{subequations}
\label{eq: noises appendix}
\begin{widetext}
\begin{equation}
    \begin{split}
    \langle\zeta_{i}\!\left(t\right)&F_{a,pq}\!\left(t'\right)\rangle=-\frac{i}{\hbar}\frac{\partial^2 H}{\partial \alpha_{i}^\dag\partial C^\dag_{a,p}}C_{a,q}^\dag(t)\delta(t-t')=-\frac{i}{\hbar}\sum_r\frac{\partial^2 H}{\partial \alpha_{i}^\dag\partial \rho_{a',rp}}\rho_{a',rq}(t)\delta(t-t'),\\
    \langle\zeta^\dag_{i}\!\left(t\right)&F_{a,pq}^\dag\!\left(t'\right)\rangle=\frac{i}{\hbar}\frac{\partial^2 H}{\partial \alpha_{i}\partial C_{a,q}}C_{a,p}(t)\delta(t-t')=\frac{i}{\hbar}\sum_r\,\rho_{a,pr}(t)\frac{\partial^2 H}{\partial \alpha_{i}\partial \rho_{a,qr}}\delta(t-t').
    \end{split}
\end{equation}
Here, in the end, we use the differentiation chain rule and substitution (\ref{eq: effective density matrix}). A similar result can be obtained for the correlators between $F_{a,pq}(t)$ and $F_{a,pq}^\dag(t)$ themselves
\begin{equation}
    \begin{split}
        \langle F_{a,pq}\!\left(t\right)F_{a',p'q'}\!\left(t'\right)\rangle=&\!-\!\frac{i}{\hbar}\sum_{r,s}\frac{\partial^2 H}{\partial \rho_{a,rp}\partial \rho_{a',sp'}}\rho_{a,rq}(t) \rho_{a',sq'}(t)\delta(t-t'),\\
        \langle F_{a,pq}^\dag\!\left(t\right)F_{a',p'q'}^\dag\!\left(t'\right)\rangle=\,&\,\frac{i}{\hbar}\,\sum_{r,s}\rho_{a,pr}(t) \rho_{a',p's}(t)\frac{\partial^2 H}{\partial \rho_{a,qr}\partial \rho_{a',q's}}\delta(t-t').\!\!\!
    \end{split}
\end{equation}
\end{widetext}
All the remaining correlators are trivial and equal to zero
\begin{multline}
        \langle F_{a,pq}\!\left(t\right)F^\dag_{a',p'q'}\!\left(t'\right)\rangle=\langle\zeta_{i}\!\left(t\right)F^\dag_{a',p'q'}\!\left(t'\right)\rangle\\=\langle\zeta_{i}^\dag\!\left(t\right)F_{a',p'q'}\!\left(t'\right)\rangle=0.
\end{multline}
\end{subequations}

Note, that in the main text of the article, equation for the density matrix (\ref{eq: stochastic equation of motion for the effective density matrix}) specifies the form for the noise terms $F_{a,pq}^\dag(t)$ and $F_{a,pq}(t)$
\begin{equation}
\begin{gathered}
        F_{a,pq}^\dag(t)=\sum_r \rho_{a,pr}(t)\xi_{a,rq}^\dag\!\left(t\right),\\ F_{a,pq}(t)=\sum_r\xi_{a,pr}\!\left(t\right)\rho_{a,rq}(t),
\end{gathered}
\end{equation}
which introduces another pair of the noise terms $\xi_{a,pq}^\dag\!\left(t\right)$ and $\xi_{a,pq}\!\left(t\right)$. In section (\ref{subsec: stochastic equations for rho}), we state their correlation properties (\ref{eq: correlation properties for the density matrix}). One can check by the substitution that these correlators (\ref{eq: correlation properties for the density matrix}) fully reproduce the properties of $F_{a,pq}^\dag(t)$ and $F_{a,pq}(t)$ (\ref{eq: noises appendix}).

\section{Interaction with bath variables}
\label{app: incoherences}

To introduce a bath to the stochastic formalism, one could replace the initial master equation (\ref{eq: initial master equation}) with (\ref{eq: master equation with dicoherence}). An alternative way is to introduce the explicit interaction with a zero-temperature boson bath, represented by operators $\hat{b}_i(\omega)$ and $\hat{b}_i\dag(\omega)$ in the Hamiltonian
\begin{equation}
\label{eq: updated Hamiltonian}
    \hat{H}=...+\sum_i\hat{L}_i\int \hat{b}_i^\dag(\omega)\frac{\hbar d\omega}{\sqrt{2\pi}}+\text{h.c.}.
\end{equation}
Here, the bath has a flat spectral density around the frequency resonant to the jump operators. After applying the Born-Markov approximation and tracing with respect to the bath variables, this additional contribution to the Hamiltonian results in the additional Lindbladian terms in (\ref{eq: master equation with dicoherence}) \cite{ressayre1978markovian, Banfi1975, agarwal1974quantum, gross1982superradiance, carmichael1999statistical}. We postpone the average over the bath's degrees of freedom and formulate the stochastic differential equations based on the updated Hamiltonian (\ref{eq: updated Hamiltonian}). 
To integrate out the stochastic variables related to the bath bosonic operators $\hat{b}_i(\omega)$ and $\hat{b}_i^\dag(\omega)$, we apply the Born-Markov approximation.

We start the derivations with  equation (\ref{eq: master equation with incoherences}). In order to get the corresponding Hamiltonian
 (\ref{eq: updated Hamiltonian}), we have to transform equation (\ref{eq: master equation with incoherences}) into equation (\ref{eq: master equation with dicoherence}), namely, to find the corresponding $\hat{L}_{i}$. Since $\Gamma_{pqrs}=\Gamma_{rspq}^*$, we can formally decompose it into a sum of smaller matrices and consequently suggest the following form for $\hat{L}_{i}$ 
\begin{equation}
\label{eq: L operators}
    \Gamma_{pqrs}=2\pi\sum_i\gamma^{(i)}_{pq}\gamma^{(i)*}_{rs},\quad \hat{L}_{i}=i\sqrt{2\pi}\sum_{p,q}\gamma^{(i)}_{pq}\hat{\sigma}_{pq}.
\end{equation}
Now, we introduce a simple model of a boson reservoir 
\begin{equation*}
\begin{split}
    \hat{H}_r=&\int \hbar \omega \hat{b}^\dag\!\!\left(\omega\right)\hat{b}\!\left(\omega\right) d\omega,\\
\Big[\hat{b}\!\left(\omega\right)&,\hat{b}^\dag\!\!\left(\omega'\right)\Big]=\delta\!\left(\omega-\omega'\right),
\end{split}
\end{equation*}
which is linearly coupled to one of the $\hat{L}$-operators
\begin{equation*}
    \hat{V}_r= i\hbar\sum_{p,q}\gamma_{pq}^{(i)}\hat{\sigma}_{pq}\int \hat{b}_i^\dag(\omega)d\omega+ \text{h.c.},
\end{equation*}
Since the coupling is linear, no particular effort is needed to generalize the resulting equations to the case of all $\hat{L}$-operators and $\Gamma_{pqrs}$.
In the probabilistic language, we write
\begin{equation}
\label{eq: additional terms in the Hamiltonian}
\begin{split}
    H_r\!\left(t\right)=&\int \hbar \omega \beta^\dag\!\!\left(\omega,t\right)\beta\!\left(\omega,t\right) d\omega, \\
    V_r\!\left(t\right)=&\, i\hbar\sum_{p,q}\gamma_{qp}^{(i)}\rho_{pq}\!\left(t\right)\int \beta^\dag\!\left(\omega,t\right)d\omega\,\\-&\,i\hbar \sum_{p,q}\gamma_{pq}^{(i)*}\rho_{pq}\!\left(t\right)\int\beta\!\left(\omega,t\right)d\omega,
\end{split}
\end{equation}
where we replace the reservoir boson operators with their associated variables $\beta\!\left(\omega,t\right)$. Using equation (\ref{eq: equations of motion for the many field variables}), one gets the equations of motion for the new reservoir variables 
\begin{equation}
\label{equation: equations for the reservoir variables}
    \begin{split}
        \!\!\frac{d\beta\!\left(\omega,t\right)}{dt}=-i\omega\beta\!\left(\omega,t\right)&+\sum_{p,q}\gamma_{qp}^{(i)}\rho_{pq}\!\left(t\right)+\tilde{\zeta}\!\left(\omega,t\right),\!\!\\
        \!\!\frac{d\beta^\dag\!\left(\omega,t\right)}{dt}=i\omega\beta^\dag\!\left(\omega,t\right)&+ \sum_{p,q}\gamma_{pq}^{(i)*}\rho_{pq}\!\left(t\right)+\tilde{\zeta}^\dag\!\left(\omega,t\right),\!\!
    \end{split}
\end{equation}
where $\tilde{\zeta}\!\left(\omega,t\right)$, $\tilde{\zeta}^\dag\!\left(\omega,t\right)$ are the noise terms for the bath variables. We use tilde to distinguish them from the other noise terms. Their correlation properties are discussed later. We assume the vacuum state as initial state of the reservoir: 
\begin{equation*}
    \beta^\dag\!\!\left(\omega,0\right)=0,\quad \beta\!\left(\omega',0\right)=0,
\end{equation*}
Let us formally integrate the equations for the bath variables (\ref{equation: equations for the reservoir variables}). Note, that we are concerned with the properties of the particles and consider the bath only as an auxiliary tool. Consequently, we can neglect the atomic variables in equations (\ref{equation: equations for the reservoir variables}) because the bath variables are always multiplied by the effective density matrix (\ref{eq: additional terms in the Hamiltonian}), which leads to the product of two $\eta$-numbers disappearing after averaging. Eventually, the effect of the bath variables on the particles is described by the following stochastic processes
\begin{equation}
\label{eq: for the bath variables}
\begin{split}
    \beta\!\left(\omega,t\right)=&\int_0^td\tau \tilde{\zeta}(\omega,t-\tau)e^{-i\omega \tau},\\
    \beta^\dag\!\left(\omega,t\right)=&\int_0^td\tau \tilde{\zeta}^\dag(\omega,t-\tau)e^{i\omega \tau}.
\end{split}
\end{equation}
\begin{widetext}
Let us now update the equations for the effective density matrix. The additional terms responsible for the interaction with the bath have the following form
\begin{multline}
\label{eq: atoms bath}
        \left(\frac{d\rho_{pq}\!\left(t\right)}{dt}\right)_{\text{bath}}\!\!\!\!\!=\sum_{r}\left(\gamma_{pr}^{(i)}\rho_{rq}(t)-\rho_{pr}(t)\gamma_{rq}^{(i)}\right)\int  \beta^\dag(\omega,t)d\omega\\+\sum_{r}\left(\rho_{pr}(t)\gamma_{qr}^{(i)*}-\gamma_{rp}^{(i)*}\rho_{rq}(t)\right)\int  \beta(\omega,t)d\omega+\sum_r\left( \rho_{pr}(t)\tilde{\xi}_{rq}^\dag\!\left(t\right)+\tilde{\xi}_{pr}\!\left(t\right)\rho_{rq}(t)\right),
\end{multline}
\end{widetext}
where we introduced noise terms  for the atomic variables $\tilde{\xi}_{pq}^\dag\!\left(t\right)$, $\tilde{\xi}_{pq}\!\left(t\right)$ coupled with $\tilde{\zeta}(\omega,t)$ and $\tilde{\zeta}^\dag(\omega,t)$ through the following correlation properties
\begin{equation}
\label{eq: correlators app}
    \begin{split}
    \langle\tilde{\zeta}\!\left(\omega,t\right)\tilde{\xi}_{pq}\!\left(t'\right)\rangle&=\gamma_{pq}^{(i)}\delta(t-t'),\\
    \langle\tilde{\zeta}^\dag\!\left(\omega,t\right)\tilde{\xi}_{pq}^\dag\!\left(t'\right)\rangle&=\gamma_{qp}^{(i)*}\delta(t-t').
    \end{split}
\end{equation}
The other correlators equal to zero. Note that the right hand side does not depend on $\omega$.

The integrals in (\ref{eq: atoms bath}) together with the expressions for the bath variables (\ref{eq: for the bath variables}) give birth to the following stochastic processes
\begin{equation}
\label{eq: definition of F}
\begin{split}
        \int_{-\infty}^\infty  \beta(\omega,t)d\omega=&F(t),\\
        \int_{-\infty}^\infty  \beta^\dag(\omega,t)d\omega=&F^\dag(t).
\end{split}
\end{equation}
Their first moments are zero. Assuming equations (\ref{eq: correlators app}) and (\ref{eq: definition of F}), careful integration leads to the following non-zero correlators of the second order
\begin{equation}
\label{eq: correlators app 2}
    \begin{split}
    \langle F\!\left(t\right)\tilde{\xi}_{pq}\!\left(t'\right)\rangle&=\pi\gamma_{qp}^{(i)}\delta(t-t'-0),\\
    \langle F^\dag\!\left(t\right)\tilde{\xi}_{pq}^\dag\!\left(t'\right)\rangle&=\pi\gamma_{pq}^{(i)*}\delta(t-t'-0).
    \end{split}
\end{equation}
Since definition (\ref{eq: for the bath variables}) contains integrals over the past, the noise terms produce memory effect. In the correlation properties, it is manifested in the form of a small, but finite time shift $-0$ inside the delta-functions.

The correlation properties $(\ref{eq: correlators app 2})$ allow a simple representation in terms of independent normalized Gaussian white noise terms $f(t)$ and $f^\dag(t)$
\begin{equation}
\label{correlation properties of vecorial noise terms}
    \begin{split}
    \langle f^*\!\left(t\right)\otimes f^*\!\left(t'\right) \rangle = \langle f\!\left(t\right) \otimes f\!\left(t'\right) \rangle = 0, \\
   \langle f\!\left(t\right) \otimes f^*\!\left(t'\right) \rangle = \delta\!\left(t-t'\right),\quad
    \end{split}
\end{equation}
Same properties hold for $f^\dag(t)$. Using the introduced noise terms, we can write
\begin{equation*}
\begin{gathered}
    F\!\left(t\right)=f(t-0), \quad F^\dag\!\left(t\right)=f^\dag(t-0),\\ \tilde{\xi}_{pq}\!\left(t\right)=\pi\gamma_{pq}^{(i)}f^*(t),\quad \tilde{\xi}_{pq}^\dag\!\left(t\right)=\pi\gamma_{qp}^{(i)*}f^{\dag*}(t).
\end{gathered}
\end{equation*}
One can check that the correlations (\ref{eq: correlators app 2}) are recovered and the other correlators are zero.

In terms of the introduced noise terms, equation (\ref{eq: atoms bath}) takes the following form
\begin{widetext}
\begin{multline}
\label{eq: atoms bath 2}
        \left(\frac{d\rho_{pq}\!\left(t\right)}{dt}\right)_{\text{bath}}\!\!\!\!\!=\sum_{r}\left(\gamma_{pr}^{(i)}\rho_{rq}(t)-\rho_{pr}(t)\gamma_{rq}^{(i)}\right)f^\dag(t-0)\\+\sum_{r}\left(\rho_{pr}(t)\gamma_{qr}^{(i)*}-\gamma_{rp}^{(i)*}\rho_{rq}(t)\right)f(t-0)+\pi\sum_r\left( \rho_{pr}(t)\gamma_{qr}^{(i)*}f^{\dag*}(t)+\gamma_{pr}^{(i)}\rho_{rq}(t)f^*(t)\right).
\end{multline}
Due to the memory effect, the terms with $f(t-0)$ and $f^\dag(t-0)$ have non-zero averages, which can be interpreted as a hidden drift
\begin{equation*}
    \Bigg\langle\left(\frac{d\rho_{pq}\!\left(t\right)}{dt}\right)_{\text{bath}}\Bigg\rangle=\!\Big\langle\pi\sum_{r,s}\left(2\gamma_{pr}^{(i)}\rho_{rs}(t)\gamma_{qs}^{(i)*}-\gamma_{rq}^{(i)}\rho_{ps}(t)\gamma_{rs}^{(i)*}-\gamma_{rs}^{(i)}\rho_{sq}(t)\gamma_{rp}^{(i)*}\right)\Big\rangle.
\end{equation*}
We would like to drop $-0$ in equation (\ref{eq: atoms bath 2}) and get the equations in Ito form.  According to the derivations in Appendix \ref{app: shift}, it is possible if one adds the lost drift explicitly
\begin{multline*}
    \left(\frac{d\rho_{pq}\!\left(t\right)}{dt}\right)_{\text{bath}}\!\!\!\!\!\!\!\!\!\!\!\!\!\\=\!\pi\sum_{r,s}\left(2\gamma_{pr}^{(i)}\rho_{rs}(t)\gamma_{qs}^{(i)*}-\gamma_{rq}^{(i)}\rho_{ps}(t)\gamma_{rs}^{(i)*}-\gamma_{rs}^{(i)}\rho_{sq}(t)\gamma_{rp}^{(i)*}\right)+\sum_{r}\left(\gamma_{pr}^{(i)}\rho_{rq}(t)-\rho_{pr}(t)\gamma_{rq}^{(i)}\right)f^\dag(t)\\+\sum_{r}\left(\rho_{pr}(t)\gamma_{qr}^{(i)*}-\gamma_{rp}^{(i)*}\rho_{rq}(t)\right)f(t)+\pi\sum_r\left( \rho_{pr}(t)\gamma_{qr}^{(i)*}f^{\dag*}(t)+\gamma_{pr}^{(i)}\rho_{rq}(t)f^*(t)\right).
\end{multline*}
\end{widetext}

All the noise terms are treated now in the Ito sense. Note that the noise terms $f(t)$, $f^*(t)$, $f^\dag(t)$ and $f^{\dag*}(t)$ are multiplied by $\rho_{pq}(t)$. Since, the effective density matrix contains $\eta$-numbers, the resulting noise terms effectively do not have any impact to the evolution of the stochastic variables and can be omited. Indeed, the first-moments equal to zero and the higher moments are proportional to powers of same $\eta$-numbers, which also leads to zero. 

Now, let us include the effect of the other $L$-operators and bring back $\Gamma_{pqrs}$ (\ref{eq: L operators}) 
\begin{multline*}
    \!\!\!\!\!\!\!\!\left(\frac{d\rho_{pq}\!\left(t\right)}{dt}\right)_{\text{bath}}\\=\!\frac{1}{2}\sum_{r,s}\left(2\Gamma_{prqs}\rho_{rs}(t)-\Gamma_{rqrs}\rho_{ps}(t)-\Gamma_{rsrp}\rho_{sq}(t)\right).
\end{multline*}
If we include all the other interactions, we come to equation (\ref{eq: rate equations 2}).

\section{Shifting the noise terms}
\label{app: shift}
Recall that Stratonovich stochastic equations can be recast to Ito form \cite{gardiner2004handbook}. In this section, we perform similar derivations for replacing shifted noise term $f(t-0)$ with usual $f(t)$. Consider the following stochastic equation
\begin{equation}
\label{eq: app initial}
    \frac{dx(t)}{dt}=...+B(t)f(t-0),
\end{equation}
where $B(t)$ has its own stochastic equation
\begin{equation}
\label{eq: app equation for B}
    \frac{dB(t)}{dt}=...+D(t)f^*(t).
\end{equation}
We would like to shift the argument of the noise term and transform (\ref{eq: app initial}) into 
\begin{equation}
\label{eq: app final}
    \frac{dx(t)}{dt}=...+A(t)+B(t)f(t).
\end{equation}
Although one could just write 
\begin{equation*}
    A(t)=B(t)\left(f(t-0)-f(t)\right),
\end{equation*}
we can show that it makes a completely deterministic contribution to the equations. The easiest way to do it is to define a grid with a small time step $\Delta t$, discretize the noise terms and examine the integrated contribution of $A(t)$ to $x(t)$.  Now, for every instant of time $t_n=n\Delta t$, we introduce a set of independent complex random numbers $\varepsilon_n$ with normal distribution
\begin{equation*}
    \langle \varepsilon_n\rangle = 0, \quad \langle \varepsilon_n \varepsilon_m\rangle = \langle \varepsilon_n^* \varepsilon_m^*\rangle = 0, \quad \langle \varepsilon_n^* \varepsilon_m\rangle =\delta_{nm},
\end{equation*}
and define the noise terms through them 
\begin{equation*}
    f(t_n)=\varepsilon_n/\sqrt{\Delta t}, \quad f(t_n-0)=\varepsilon_{n-1}/\sqrt{\Delta t}.
\end{equation*}
Consequently, one gets
\begin{equation*}
    A(t_n)=B(t_n)\left(\varepsilon_{n-1}-\varepsilon_n\right)/\sqrt{\Delta t}.
\end{equation*}
Let us find its integrated contribution over a span of time between arbitrary $t$ and $t+\Delta T$, that covers several time steps $\Delta t$. Although, we assume $\Delta T$ to be small, $\Delta T$ itself is divided into $N$ even smaller time steps $\Delta t$. Consequently, we assume $N\gg 1$. We approximate the integral of $A(t)$ with a sum
\begin{multline*}
\label{eq: app integral}
    \int_{t}^{t+\Delta T}A(t')dt'\approx\sum_{m=n}^{n+N}B(t_m)\left(\varepsilon_{m-1}-\varepsilon_m\right)\sqrt{\Delta t}\\\approx\sum_{m=n}^{n+N} \underbrace{\left[B(t_{m+1})-B(t_m)\right]}_{\Delta B(t_{m})}\varepsilon_m\sqrt{\Delta t}.
\end{multline*}
In the transition to the last line, we have changed the limits of summation and neglected the edge terms. Here, $\Delta B(t_{m})$ must contain deterministic and noisy components proportional to $\Delta t$ and $\sqrt{\Delta t}$ respectively. The deterministic component will make the integrand proportional to $\Delta t^{3/2}$, which is too small to give a finite result after integration. The noisy component leads to an integrand proportional to $\Delta t$, which can give a finite value in the limit $\Delta t \to 0$. Recalling equation (\ref{eq: app equation for B}), we can write
\begin{multline*}
    \int_{t}^{t+\Delta T}A(t')dt'\approx\sum_{m=n}^{n+N} D(t_m)\varepsilon_m\varepsilon_m^*\Delta t\\\approx D(t)\sum_{m=n}^{n+N} \varepsilon_m\varepsilon_m^*\Delta t.
\end{multline*}
Here, we can approximate $t_m$ with $t$, since the difference creates contributions of higher orders with respect to $\Delta T$. 

The sum $\sum_{m=n}^{n+N} \varepsilon_m\varepsilon_m^*\Delta t$ in the resulting expression can be seen as a random variable, so let us explore its properties. Its average is equal to $\Delta T$ and is independent from $N$. However, the variance equals to $N\Delta t^2=\Delta T^2/N$. In the limit of $N\to \infty$, the variance tends to zero, and $\sum_{m=n}^{n+N} \varepsilon_m\varepsilon_m^*\Delta t$ behaves as an ordinary number equal to $\Delta T$. Assuming that $\Delta T$ is small, we conclude 
\begin{equation}
    A(t)=D(t).
\end{equation}
On the other hand, the direct use of correlation properties leads to
\begin{equation*}
    D(t)=\langle B(t)f(t-0) \rangle_t,
\end{equation*}
where index $_t$ means that we average over the noise terms in the vicinity of $t$.

The result can be formulated in an elegant way. Due to memory effect, the noise term $B(t)f(t-0)$ in (\ref{eq: app initial}) has a non-zero average value, which can be seen as a hidden contribution to the total drift. If we omit $-0$, the Ito interpretation leads to zero average. The additional drift $A(t)$ in equation (\ref{eq: app final}) compensates this loss
\begin{equation*}
\label{eq: app solution}
    A(t)=\langle B(t)f(t-0) \rangle_t,
\end{equation*}
which makes it similar to the transition from Stratonovitch to Ito form of stochastic equations. 

\section{Drift gauge}\label{app: drift gauge}
To give an example of how the concept of stochastic gauge transformations from section \ref{sec: stochastic gauges} can be applied, we derive the so-called drift gauge \cite{2005'DeuarPhD} in the spirit of Girsanov's theorem \cite{liptser2001statistics}. In certain cases, stochastic differential equations
\begin{equation*}
    \frac{d\textbf{x}(t)}{dt}=\textbf{A}(\textbf{x}(t),t)+\bm{\xi}(\textbf{x}(t),t)
\end{equation*}
possesses drift terms $\textbf{A}(\textbf{x},t)$ that force the stochastic trajectories to diverge. Here, $\bm{\xi}(\textbf{x},t)$ are Gaussian white noise terms with zero first moments and arbitrary second order correlators. Unfortunately, neglecting divergent trajectories leads to wrong expectation values. To resolve problems related to the numerical instability of diverging trajectories, one can choose other stochastic equations $\textbf{x}'(t)$ that have different drift terms $\textbf{A}'(\textbf{x}',t)$
\begin{equation*}
    \frac{d\textbf{x}'}{dt}=\textbf{A}'(\textbf{x}',t)+\bm{\xi}(\textbf{x}'(t),t)
\end{equation*}
and have the same initial conditions. To compensate this change, one can introduce weight coefficient $\Omega(t)=e^{C_0(t)}$ that can be used to calculate expectation values based on the new stochastic variables 
\begin{equation*}
    \langle f(\textbf{x}(t))\rangle=\langle f(\textbf{x}'(t))\Omega(t)\rangle.
\end{equation*}
Consequently, the new characteristic function has the following form
\begin{equation*}
    \chi'(\bm{\lambda},t)=\Big\langle \exp\!{\big(\bm{\lambda}\cdot\textbf{x}'(t)+C_0\!\left(t\right)\!\big)}\Big\rangle.
\end{equation*}
At the level of the $P$ distribution, it corresponds to the following expression
\begin{equation*}
    P(t,\textbf{x})=\Big\langle\delta(\textbf{x}(t)-\textbf{x})\Omega(t)\Big\rangle,
\end{equation*}
which shows, that $\Omega(t)$ is indeed responsible for weighting different statistical trajectories.

We can always formally write an equation of motion for $C_0(t)$
\begin{equation*}
    \frac{dC_0(t)}{dt}=A_0(\textbf{x}'(t),t)+\xi_0(\textbf{x}'(t),t),
\end{equation*}
where $A_0$ is a new drift and $\xi_0$ is a new Gaussian white noise term. We assume that $\xi_0(t)$ has zero average and yet unknown correlation properties
\begin{equation*}
\begin{gathered}
    \langle \xi_0(\textbf{x},t)\xi_0(\textbf{x},t')\rangle=\sigma_{0}(\textbf{x},t)\delta(t-t'), \\
    \langle \bm{\xi}(\textbf{x},t)\xi_0(\textbf{x},t')\rangle=\bm{\sigma}(\textbf{x},t)\delta(t-t').
\end{gathered}
\end{equation*}
The main goal is to find drift $A_0$ for the weight coefficient and correlation properties $\sigma_{0}$ and $\bm{\sigma}$, that compensate the change of drift terms $\Delta \textbf{A}=\textbf{A}'-\textbf{A}$ at the level of the characteristic function. Acting in the spirit of mathematical induction, we assume that $\chi'\!\left(\bm{\lambda},t\right)=\chi\!\left(\bm{\lambda},t\right)$. Let us see, if the same happens to the derivatives 
\begin{multline*}
    \frac{\partial}{\partial t}\left[\chi'\!\left(\bm{\lambda},t\right)-\chi\!\left(\bm{\lambda},t\right)\right]\\=\Bigg(\bm{\lambda}\cdot \left[ \Delta\textbf{A}\left(\frac{\partial}{\partial \bm{\lambda}},t\right)+\bm{\sigma}\left(\frac{\partial}{\partial \bm{\lambda}},t\right)\right]\\A_0\left(\frac{\partial}{\partial \bm{\lambda}},t\right)+\frac{1}{2}\sigma_{0}\left(\frac{\partial}{\partial \bm{\lambda}},t\right)\Bigg)\chi\!\left(\bm{\lambda},t\right).
\end{multline*}
In derivation of this expression we have used the Ito's lemma discussed in Appendix \ref{app: ito}. To make the right hand side equal to zero for any $\bm{\lambda}$, we have to choose the following correlation properties for the noise terms
\begin{equation*}
\begin{gathered}
    \langle \xi_0(\textbf{x}',t)\xi_0(\textbf{x}',t')\rangle=-2A_0(\textbf{x}',t)\delta(t-t'), \\
    \langle \bm{\xi}(\textbf{x}',t)\xi_0(\textbf{x}',t')\rangle=-\Delta\textbf{A}(\textbf{x}',t)\delta(t-t').
\end{gathered}
\end{equation*}
Which is the main content of the drift gauge. The main feature of our derivations is that we do not base them on the properties of $\hat{\Lambda}$-projectors: our result is applicable to any system of stochastic trajectories including the equations of motion for the effective density matrix.

\end{document}